\documentclass[preprint2]{aastex}

\shorttitle{Quasar Variability in the P-Q Survey}
\shortauthors{Bauer et al.}

\begin{document}

\begin{titlepage}

\title{Quasar Optical Variability in the Palomar-QUEST Survey}
\author{Anne Bauer\altaffilmark{1,2}, Charles Baltay\altaffilmark{1}, Paolo Coppi\altaffilmark{1}, Nancy Ellman\altaffilmark{1}, Jonathan Jerke\altaffilmark{1}, David Rabinowitz\altaffilmark{1}, Richard Scalzo\altaffilmark{1}}
\email{anne.bauer@aya.yale.edu}
\altaffiltext{1}{Yale University, Department of Physics, P.O. Box 208120, New Haven, CT 06520-8120, USA}
\altaffiltext{2}{Universit\"{a}ts-Sternwarte M\"{u}nchen, Scheinerstr. 1, D-81679 M\"{u}nchen, Germany}

\begin{abstract}

The ensemble variability properties of nearly 23,000 quasars are studied 
using the Palomar-QUEST Survey.  The survey has covered 15,000 square 
degrees multiple times over 3.5 years using 7 optical filters, and has 
been calibrated specifically for variability work.  Palomar-QUEST 
allows for the study of rare objects using multiple 
epochs of consistently calibrated, homogeneous data, obviating the common 
problem of generating comparable measurements from disparate datasets.  
A power law fit to the quasar structure function versus time yields an 
index of 0.432 $\pm$ 0.024 for our best measured sample.  We see the 
commonly reported anticorrelation between average optical variability 
amplitude and optical luminosity, and measure the logarithmic decrease in 
variability amplitude to scale as 
the logarithm of the luminosity times 0.205 $\pm$ 0.002.  Black hole mass is 
positively correlated with variability amplitude over three orders of 
magnitude in mass.  Quasar variability amplitude is seen to decrease with 
Eddington ratio as a step function with a transition around Eddington ratio of 
0.5.  The higher variability at low Eddington ratios is due to 
excess power at timescales shorter than roughly 300 days.  X-ray and radio 
measurements exist for subsets of the quasar sample.  We observe an 
anticorrelation between optical variability amplitude and X-ray luminosity.  
No significant correlation is seen between average optical 
variability properties and radio luminosity.  The timescales of quasar 
fluctuations are suggestive of accretion disk instabilities.  The 
relationships seen between variability, Eddington ratio, and radio and 
X-ray emission are discussed in terms of a possible link between the behavior 
of quasars and black hole X-ray binaries.

\end{abstract}


\maketitle

\end{titlepage}

\section{Introduction}

Quasars are known to be variable in the optical as well as 
other wavelengths.  However, the physics behind the fluctuations is 
not understood.  Flares due to accretion disk instabilities are a 
promising mechanism (e.g. \cite{pereyra06}), but other possible sources of 
variability include starbursts in the host galaxies (e.g. \cite{aretxaga97}) 
or microlensing of the quasars (e.g. \cite{zackrisson03}).  
Each candidate model can predict variability 
characteristics such as time dependences and trends with quasar luminosity.  
Few theoretical results are available with which to compare 
(e.g. \cite{kawaguchi98}; \cite{hawkins02}), yet qualitative relationships 
between variability amplitude and quasar properties can elucidate how 
the fluctuation mechanisms relate to different physical regions and 
processes in AGN.  For example, consider black hole X-ray binaries 
(BHXBs), which are much less massive systems that 
also contain central black holes, accretion disks, and sometimes radio 
jets.  Correlations between the variability and other properties such as 
radio emission and Eddington ratio in BHXBs yield insight into the 
workings of the system as a whole.  
%
%These binaries are seen to inhabit three typical states, known 
%as low/hard, high/soft, and intermediate (or very high), 
%which are dominated to differing extents by accretion disk and 
%coronal emission in the X-rays, and jet emission in radio wavelengths.  
%There may not be a direct connection between the 
%physics of black hole X-ray binaries and quasars, since the complicated 
%processes that take place in an accretion disk may not scale simply with 
%mass.  However, they are thought to have similar underlying geometry, 
%and recent studies have found some common emission 
%characteristics, particularly in X-ray wavelengths (see section 
%\ref{bhxb_section} for further discussion).  
The ability to characterize the BHXB behavior 
into discrete states using 
%parameters such as variability, Eddington 
%ratio, and radio emission 
these properties 
is intriguing in terms of the study of quasars.

Recently, as large scale surveys have become feasible, the ensemble 
optical variability of thousands of quasars has been studied 
over timescales up to a few years (e.g. \cite{vandenberk04}; 
\cite{adam}; \cite{wilhite08}) 
as well as, using archival photographic plates, several decades 
(\cite{devries05}; \cite{sesar06}).  
Most variability work on large quasar samples has been performed by 
comparing one wide-area dataset with an independent, previous one.  This 
method has yielded many valuable results, but it invariably requires 
complicated calibrations to correct for the disparate nature of the different 
measurements.  
The Palomar-QUEST Survey, by repeatedly covering 15,000 square degrees of 
sky in optical bands, allows for a robust, consistent analysis of roughly 
23,000 spectroscopically confirmed quasars on many timescales 
over a total span of 3.5 years.  We have cleaned and calibrated the 
survey data for variability purposes and have studied the 
average quasar variability amplitude with respect to fluctuation timescale, 
quasar luminosity, mass, Eddington ratio, X-ray loudness, and radio 
loudness.

This work uses typically four measurements of each quasar in the 
sample.  Our results describe only the ensemble behavior of the 23,000 
objects, as the data are not sufficient to study the detailed 
variability of each quasar.  Individual AGN have diverse lightcurve 
properties;  \cite{collier01} find a wide range of variability timescales 
in 12 well-sampled Seyfert 1 galaxies, sometimes measuring several characteristic 
timescales in a single AGN's lightcurve.  In an ensemble study such as 
our work, using a few measurements of many AGN, such diversity of behavior will 
be either missed completely or ignored while determining average variability 
properties.  Future work with even larger surveys may be able to bridge the 
current gap between studies with high quasar statisics and those 
with detailed lightcurves.

We describe the Palomar-QUEST Survey and its calibrations 
in \S 2.  The quasar sample is introduced in \S 3.  The structure 
function, which is the statistical quantity we use to analyze the data, 
is discussed in \S 4.  Quasar properties, such as mass 
and luminosity, are often correlated;  to study the variability dependence 
on each separately we divide the data into multi-dimensional bins.  
This procedure is described in \S 5.  \S 6 presents our results, 
which are discussed and compared to previous work in \S 7.  Conclusions 
are given in \S 8.  Appendix A describes in detail the modelling of the 
data in order to understand the effects of our observing cadence on the 
structure function analysis.

\section{The Palomar-QUEST Survey}

\subsection{Overview}

\begin{figure}
\begin{center}
\plotone{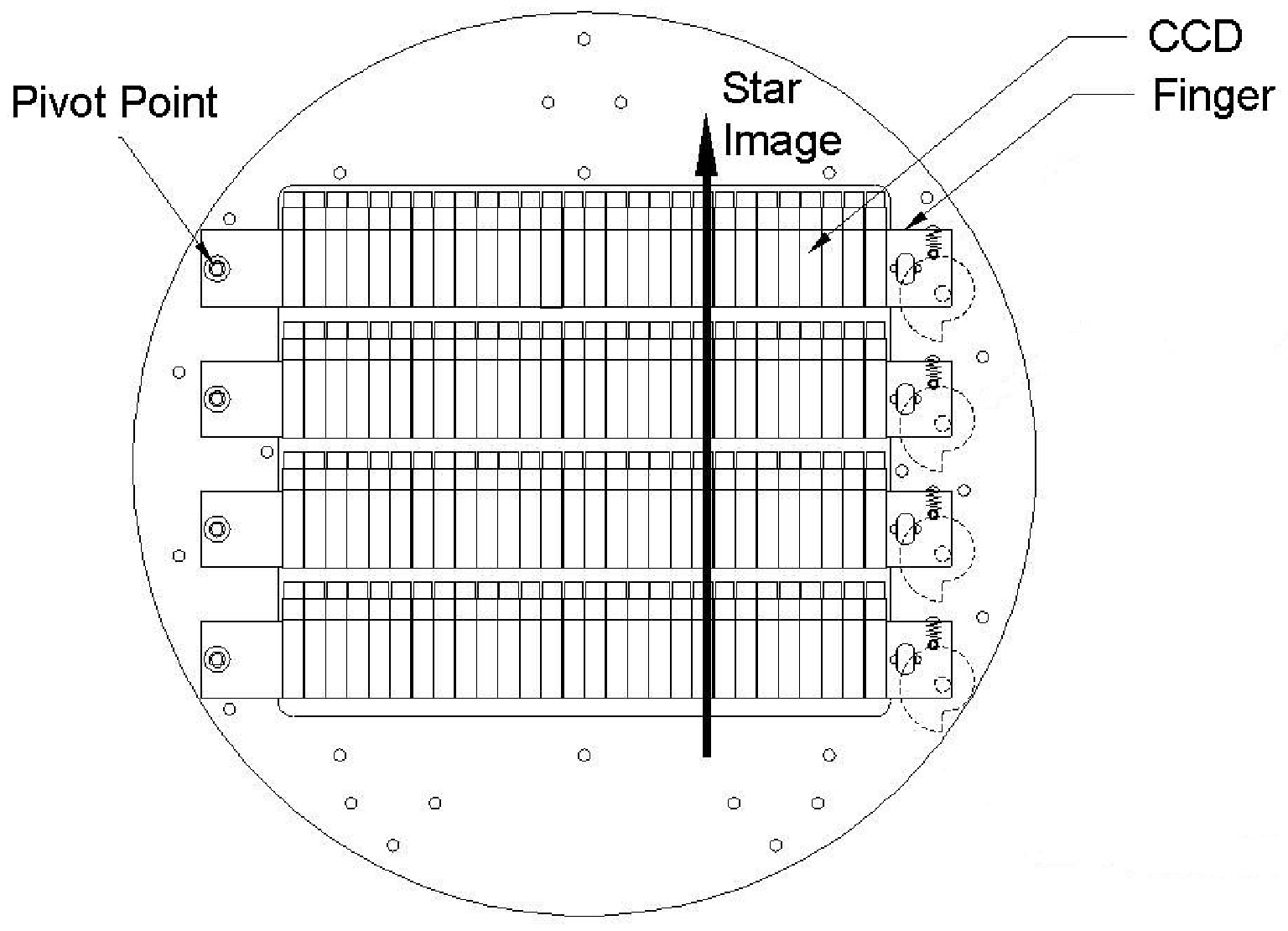}
\end{center}
\caption{The QUEST camera design, showing the 112 CCDs arranged on 4 rows, or fingers.}
\label{camera_figure}
\end{figure}

The Palomar-QUEST Survey covers roughly 15,000 square degrees 
multiple times 
with seven optical filters: Johnson UBRI and SDSS r'i'z'.  We use the 
48'' Samuel Oschin Schmidt Telescope at Palomar Observatory to scan the 
sky between declinations $\pm25^{\circ}$, excluding the galactic plane.  
The survey is 
unique in its repeated coverage of such a large area.  The time between 
different scans over the same coordinates ranges from several hours to 
three years, typically totalling 4 or 5 passes per filter.  Often the same 
sky coordinates were covered twice in one season, and the whole sky area 
was observed each year.  The wide 
coverage is made possible by the QUEST large-area CCD camera, which 
was custom built for the survey.  The camera consists of 112 CCDs, each 
600 $\times$ 2400 pixels, which are arranged in 4 rows by 28 columns.  
The survey data were taken in driftscan mode, where the telescope is kept 
in a fixed position throughout a scan and the stars drift across the 
field of view of the camera such that every star crosses each of the 4 
rows of chips.  The stars move parallel to the $y$ coordinates of the chips, 
and the CCDs are read out in synchrony with the sky motion.  The $y$ chip 
axis therefore is the RA east/west direction, as well as the axis of 
increasing time throughout the scan.  A separate filter covers each row, 
or finger, 
so that driftscanning yields nearly simultaneous observations in 4 filters 
with a fixed exposure time of roughly 140 seconds.  The geometry of the camera 
is shown in figure \ref{camera_figure}.  A typical 8 hour night of observing 
yields 500 square 
degrees of data in 4 filters.  The camera hardware is discussed in detail in 
\cite{camera_paper}.  

The software used to process the data and yield calibrated object catalogs 
was written specifically for the survey.  The object detection and primary 
photometry code fit the data to an empirical point spread function (PSF) model, 
which allows for appropriate deblending of neighboring objects.  Photometry 
is also measured using several apertures.  Astrometry is determined through 
matching with the USNO A-2.0 catalog (\cite{usno}) and is good to 0.1''.  
The QUEST processing software is described in further detail in \cite{software_paper}.

\subsection{Variability Calibration}

\subsubsection{Summary of the Standard Calibration}

The standard QUEST photometric calibration consists of two steps: 
first to correct for the different intrinsic properties of each CCD, 
and second to compensate for poor weather.  The first routine takes 
into account both the different response levels of the chips, as well as 
nonlinearities in some chips.  The correction coefficients were 
calculated using a few scans, taken under good conditions, 
which overlap with the Sloan Digital Sky Survey Data Release 
4 (\cite{sdss4}) which was used as the reference standard.  The second 
step, or extinction correction, uses as a calibration standard for 
each square degree of sky the QUEST scan over that area for which the 
objects' measurements are brightest.  We use our own data as the 
calibration standard because there is currently no survey which covers 
our entire sky area with photometry that is accurate to our precision.  
Correcting for suboptimal observing conditions is essential to the 
calibration of the QUEST Survey;  we have accumulated so much data 
because we observe on non-photometric nights.  54\% of the extinctions 
as measured with respect to the extinction correction's reference scans 
are greater than 0.1 magnitudes.  33\% are greater than 0.2 magnitudes.  
Clearly, to take advantage of the large dataset, we must use an effective 
extinction correction.  Further description of the standard calibration 
software and performance can be found in \cite{software_paper}.  
The calibration is accurate to about 4\% in the 
R, I, r', and i' filters, with roughly 7.5\% of the measurements disagreeing 
with the accepted value by more than 3$\sigma$.  These outliers 
appear variabile, although the true source of most of them is 
simply calibration error.  In order to use the Palomar-QUEST data for 
variability purposes, we have developed a new, relative calibration that 
improves both the average precision of the photometry as well as the 
percentage of outliers.

The goal of the standard calibration is to produce the most accurate 
and precise possible average flux measurements of each object.  
The goal of the relative calibration is for each measurement of an object 
to be as consistent as possible with each other measurement of the same 
object.  The accuracy of the average flux is less important than 
the ability to discern intrinsic variability.  This shift in priorities 
between the standard and relative calibrations leads to some differences 
between the two procedures.

\subsubsection{Relative Correction vs. RA or Time}

The standard calibration's extinction correction 
compares overlapping QUEST scans, assumes the one with the brightest 
measurements of common objects to be photometric, and corrects the other 
data using that scan as standard.  The correction is done using one 
multiplicative flux constant for each chip for each degree right ascension, 
which corresponds 
to about fifteen minutes of data taken in driftscan mode.  This routine 
is meant to correct for slowly changing weather which causes extinction 
in the data.  However, it is not 
suited to a relative calibration.  It is possible that our scan with 
the least extinction may in fact be slightly extincted, moreover in a 
time-dependent fashion.  When we correct other data to this scan 
we will introduce unwanted time-dependent residuals.  The resulting flux 
measurements will be as close to the photometric values as possible;  
therefore this procedure is indeed best for the standard calibration.  
However, the introduced residuals will cause inconsistencies between 
individual measurements of the objects.  

To avoid this situation 
we select as the ``best'' scan not the one with the brightest measurements, 
but the one with the most stable properties 
throughout the sky area being calibrated.  This criterion is quantified 
using the statistic $S$ given in equation \ref{stat_eq}.  
If the QUEST measurements under consideration are consistent with the 
accepted values, the $S$ distribution will be Gaussian and have unit width.
We use the average QUEST magnitudes as the accepted values, and compare them 
to the magnitudes measured on an individual QUEST scan.  If the given scan has 
changing conditions then the distribution will have greater than unit width.  
If the given scan has constant data quality then the distribution will 
be narrow, although it may not be centered on zero.  We choose, as the 
``best'' scan for relative calibration purposes, the scan which yields the 
narrowest such distribution.  The scans are ranked separately for 
each 1/16 square degree of data.  

The ``best'' scan also has the brightest object measurements in only about 
half of the cases.  
Much of the discrepancy occurs because different sets of scans are compared in the 
two cases.  The standard extinction corrections are run separately on groups 
of scans centered near typical QUEST declination pointings, and extinctions 
are calculated by only comparing data from the same chip on different scans.  
Because the data are dithered by typically one quarter 
of a chip width, this organizational scheme includes enough chip overlap to 
calibrate the vast majority of the QUEST Survey.  The extinction calculated 
for a scan using an overlapping subset of a chip are extrapolated to the 
remainder of the chip.  The relative calibration, however, compares all 
overlapping data, including that from neighboring chips or unusual pointings.  
Furthermore, often a chip will be split between two 1/4 degree declination bins.  
The relative calibration results are recalculated for each bin rather than 
extrapolated.  Of the ``best'' scans with non-zero extinction, 
about 15\% have extinctions over 0.1 magnitudes, and 
4\% over 0.2 magnitudes.  These numbers are much smaller than the total data 
fractions with these extinctions;  therefore the relative calibration favors 
low-extinction data.  However, the ``best'' scans are often indeed extincted.  

The Sloan Digital Sky Survey (SDSS) overlaps roughly one quarter of the QUEST 
sky area.  For these regions, we use SDSS data as the ``best'' available scan.

\begin{equation}
S = \frac{m_{\mathrm{accepted}} - m_{\mathrm{QUEST}}}{\sqrt{\delta m_{\mathrm{accepted}}^{2} + \delta m_{\mathrm{QUEST}}^{2}}} \label{stat_eq} 
\end{equation}

Using the ``best'' scan, we apply an RA-dependent, or equivalently a 
time-dependent, correction to all overlapping scans in the 1/16 
square degree under consideration.  In each scan, ``good quality'' objects 
are chosen with which to determine the calibration, which is subsequently 
applied to all objects.  The ``good quality'' calibration objects have 
magnitude greater than 16, so that they are not saturated, and 
error less than 0.08 magnitudes, so that they have good statistics.  
The error limit corresponds to a magnitude limit around 19.7 in the 
R, r', I, and i' filters.  Objects with neighbors within 15'' are not 
used for calibration purposes in order to avoid flux contamination 
from nearby sources.  The ``best'' scan's calibration objects are matched with 
those from an overlapping scan, and a line is fit such that 

\begin{equation}
m_{\mathrm{best}} - m_{\mathrm{overlap}} = a \times \mathrm{RA} + b.
\end{equation}

$a$ and $b$ are determined by chi squared minimization, and used to correct 
the magnitude measurements $m$ from the secondary, overlapping scan.  This process is 
repeated until the ``best'' scan is used to correct all scans that it overlaps.  
There may be a scan in the 1/16 square degree that does not overlap the 
``best'' scan;  if it does overlap a scan that has been calibrated to the ``best'' 
scan, it is calibrated to this intermediate data.  
The application of a linear fit rather than a simple constant better 
approximates changing weather conditions across the sky region.

\subsubsection{Relative Correction vs. Declination}

Variation in chip sensitivity by $x$ coordinate, which corresponds to 
the declination axis, is accounted for in the first step of the 
standard calibration.  We apply this correction as the first step of 
the relative calibration as well, since we do want to correct for 
$x$-dependent instrumental effects.  There exist, however, residual $x$ 
dependencies since some causes of the sensitivity variations change 
with time.  For example, scattered moonlight yields $x$-dependent features 
in the data, and it changes from night to night and also within a single 
scan.  The residual 
calibration errors in $x$ usually show up as discrete jumps rather than 
smooth trends;  therefore the relative calibration bins the objects 
by $x$ coordinate and corrects for any discrete jumps observed between 
objects' magnitudes as measured on an individual scan, and our best 
estimate of those magnitudes.

Because QUEST images are 600 pixels wide in the $x$ direction and we 
often dither our central scan declinations by 150 pixels or more, 
there is typically poor $x$ overlap between different scans.  It is therefore 
impractical to find a single best scan that can be used to correct the 
rest of the data.  After the $y$ dependent correction is applied to each 
scan, each object's measurements are re-averaged to find an updated mean 
magnitude.  The new mean magnitude is then used as the best value to which 
to correct.  The $x$ dependent calibration is done once for each five degrees 
RA, or 75 minutes of driftscan, in order to account for slow changes in 
conditions while accumulating enough statistics to fit a meaningful trend 
in the $x$ direction.  The correction is usually insignificant;  only 
$\sim2$\% of measurements are changed by over 0.01 magnitude.  However, 
about 0.5\% 
are corrected by more than their statistical errors, with occasional 
corrections of several tenths of a magnitude.

\subsubsection{Further Quality Cuts}

More quality cuts are made on the data during the relative calibration 
than during the standard one, for the purposes of eliminating any 
remaining calibration tails.  These cuts are quite strict;  however they 
are important to ensuring a reliable variability analysis.

Several cuts are invoked to eliminate spacially extended objects and 
artifacts from 
the dataset.  The detection and main photometry routines in the QUEST 
processing software use the data's point spread function (PSF) 
to deconvolve and measure overlapping objects.  
PSF routines, however, do not properly measure extended objects, yielding 
magnitudes that will 
depend on changing parameters like the seeing.  Because the vast majority 
of quasars appear pointlike in QUEST data, we do not implement further 
analysis techniques to handle the variability of extended sources.  
Instead, we use the PSF magnitudes for variability work and simply eliminate 
objects that appear extended.  Because artifacts like unmasked bad columns 
and saturation trails act like extended objects in the analysis, these 
are also removed at this stage.  Two cuts are implemented for this purpose; 
their effectiveness was tested using areas of overlap with SDSS.

If a detection appears extended, the PSF routine may deconvolve it 
into several distinct closeby objects.  To eliminate these from the dataset 
we remove all objects that have a neighbor within 3.5''.  The typical 
seeing on a good quality night is roughly 2.3''.  This cut throws away a 
fraction of legitimate stars, but is effective at removing 
spurious detections that are deblended from a single, extended object 
and often appear variable.  About 11\% of all detections are cut at this 
stage.  Roughly 7\% of the eliminated objects are stars.

The QUEST photometry routines measure both PSF and aperture fluxes.  
For pointlike objects, the aperture flux measured in a 1 FWHM radius 
is typically 15\% smaller than the PSF flux.  This difference is corrected 
by separate calibration terms for the PSF and aperture results, yielding 
consistent measurements for pointlike objects.  If an object 
is extended, the relationship between the PSF and aperture fluxes will 
change;  a comparison of the two fluxes is therefore a test of 
how well the object conforms to the PSF profile.  We eliminate any 
measurements for which the two measurements differ by greater than 0.1 
magnitude.  This cut eliminates about 20\% of all measurements.  However, 
only roughly 6\% of those eliminated are measurements of stars.

Measurements made in regions of the 
data with high background noise are also eliminated.  Noisy areas often 
yield poor measurements due to errors in background subtraction, 
particularly in cases where only small regions of the image have increased 
sky levels (such as areas close to saturated stars).  For each 1/16 
square degree of data, the background noise is calculated around each 
object inside an annulus with inner and outer radii of 30 and 35 pixels 
from the object's position.  The median of the region's noise distribution 
is determined, and the distribution's natural width is assumed to be 
reflected by the shape of its lower half.  Any asymmetric tail of higher 
noise will be due to areas with unusually high background, which are likely 
to yield bad measurements.  All objects with 
\begin{equation}
\mathrm{noise} > \mathrm{\overline{noise}} + (\mathrm{\overline{noise}} - \mathrm{noise_{min}})
\end{equation}
are discarded, where $\mathrm{\overline{noise}}$ is the median background 
noise value and $\mathrm{noise_{min}}$ is the smallest measured noise.  
This eliminates roughly 7\% of the data.

Furthermore, for 
each 1/16 square degree, any scan that contains more than 15\% of its 
measurements outside $\pm 3$ in the distribution of $S$ as defined in 
equation \ref{stat_eq} 
is disregarded.  As before, $m_{accepted}$ is taken to be the average 
QUEST magnitude measurement of an object and $m_{QUEST}$ is the magnitude 
measurement from the scan in question.  This cut eliminates 
about 3\% of the data, which 
have serious calibration problems that have not been specifically 
anticipated by the code.

\subsubsection{Evaluation of the Relative Calibration}

\begin{figure}
\begin{center}
\plotone{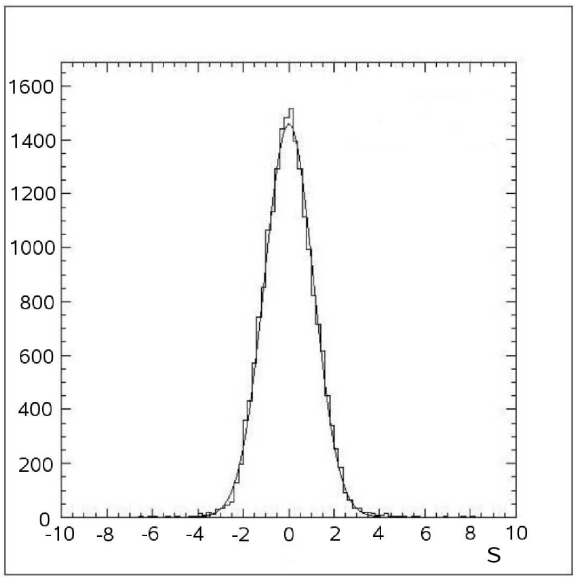}
\end{center}
\caption{Distribution of $S$, as defined in equation \ref{stat_eq}, after relative calibration.  A Gaussian fit is superimposed.}
\label{rel_stat}
\end{figure}

It is important to have as many measurements of an object as possible if 
one is to study its variability.  In order to accumulate a maximum 
number of comparable measurements, we have calibrated together data from 
the Johnson R and SDSS r' filters, and similarly the Johnson I and SDSS i' 
filters.  The resulting hybrid bandpasses will be referred to as Rr and Ii.  
This cross-calibration introduces some error due to uncorrected color terms, 
but greatly increases the power of the data to study variability by roughly 
doubling the number of comparable measurements of each object.  The 
color error introduced is larger for the Ii data than for the Rr, since 
the Johnson R and SDSS r' filters have very similar wavelength ranges.  
Because of this effect, the Rr data are used throughout the following 
analysis rather than the Ii data.  The systematic errors associated with the 
relative calibration for a random sample of pointlike objects are 0.7\% for 
the Rr data and 1.3\% for the Ii data.  They are significantly smaller than the 
$\sim 4$\% seen in the standard calibration.  
Figure \ref{rel_stat} shows the distribution, for a sample of bright 
(r magnitude $\sim$16) objects, of $S$ 
after the relative calibration is applied to the data.  
The relative calibration has very small non-Gaussian tails;  now 
about 2\% of the data 
in a high galactic latitude sample region lie outside $\pm$3, rather than 
7.5\%.  Furthermore, 
as the systematic errors have decreased, the statistic $S$ is sensitive to 
smaller variations.  

To estimate the Rr measurement error for quasars due to 
uncorrected color terms, we convolved the 
R and r' filter transmission curves with various template spectra.  The 
\cite{pickles98} template spectra of A through G main sequence stars are 
used to estimate differences in how the R and r' filters measure common stars.  
The calculated broadband R over r flux ratios for 
these sample objects average $0.986 \pm 0.008$, indicating that the 
two filters treat these stars similarly.  In practice, the average offset 
between the R and r fluxes of typical objects is taken out by the 
relative calibration.  The ratio of R over r calculated flux for the composite 
quasar spectrum published in \cite{compositespectrum}, measured at a range 
of quasar redshifts between 0.4 and 2.2, is slightly different at $1.023 \pm 0.008$.  
Quasars get bluer when they brighten;  \cite{wilhite05} measured 
the changes to be primarily due to the continuum rather than emission lines.  
They approximated the change in spectral shape between quasars in bright and 
faint phases as $\Delta F = (\lambda/3060 \mathrm{\AA})^{-2.00}$.  The R and r' 
filter curves yield the same broadband flux for this $\Delta F$ spectrum.  
Therefore we do not expect the quasar color change to increase our color 
terms, and we estimate the quasar color terms to be roughly 3.5\% of the flux, 
as this is the percent difference between the R to r ratios for main sequence 
stars and quasars.  
A 3.5\% flux variation corresponds to a structure function amplitude of 
$\sim 0.05$;  fits made in this work include data with average variability 
starting at roughly twice this level, and usually showing much stronger 
fluctuations.  
Therefore the color terms introduced into the quasar measurements by 
calibrating together R and r data will not strongly affect our results.

We can quantitatively compare the variability that we see with that measured 
by other surveys.  \cite{huber06}, as part of the Faint Sky Variability Survey 
(FSVS), studied $\sim$ 3.5 degrees$^{2}$ observed $12 - 15$ times over 
timespans from hours to 3 years and determined what fraction of observed 
objects appeared variable, for various object magnitude bins and variability 
amplitudes.  The relatively small areal coverage of their study means that the 
statistics are low for rare, highly variable objects;  however they see that 
roughly 0.5\% to 1\% of all objects between magnitudes 17.5 and 18.5 have 
variability amplitudes of at least 0.1 magnitude in the V band.  We see that 
0.6\% of objects with Rr magnitudes between 17.5 and 18.5 vary with amplitudes 
greater than 0.1 magnitude simultaneously in the Rr and Ii bands.  To determine 
this statistic we require a 0.1 magnitude jump to be significant to $3 \sigma$ 
in both the Rr and Ii bands.  This requirement is not satisfied by most objects 
fainter than magnitude Rr $\sim$ 18.5;  therefore we limit the comparison to 
objects brighter than this value.  In this range of magnitude and variability 
amplitude overlap between Palomar-QUEST and the FSVS, our observations agree 
very well.

\section{The Quasar Sample}

We have studied the variability of a sample of SDSS spectroscopically 
identified quasars.  SDSS covers a substantial fraction (roughly one 
quarter) of the QUEST area, and has to date measured over 100,000 quasars.  
Spectroscopic identification ensures that the sample is close to 100\% pure, 
and the spectra provide supplemental information such as line widths that 
allow us to estimate parameters like the mass of the quasar's black hole.  
Furthermore, accurate knowledge of the quasar redshifts is crucial for 
studying the rest-frame time dependence of variability.

\begin{figure}
\begin{center}
\plotone{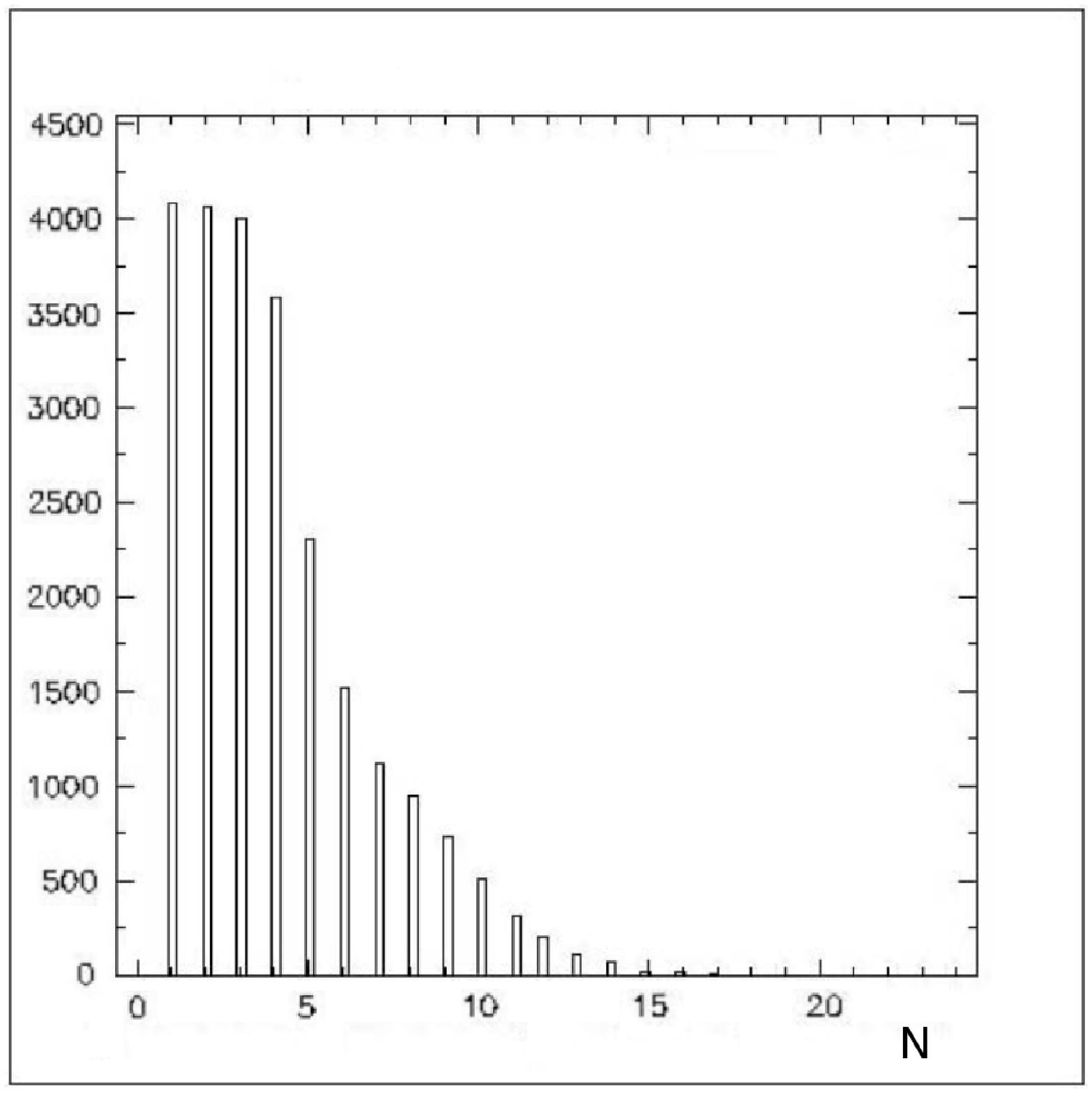}
\end{center}
\caption{Number of quasars with $N$ good Rr measurements, versus $N$.}
\label{num_hist}
\end{figure}

25,043 SDSS spectroscopic quasars, published as of April 2007, lie in 
the QUEST sky area.  On average we have slightly more than four good 
quality, relatively calibrated measurements of each quasar in the Rr 
band.  For a few objects we have over twenty;  a histogram of the number 
of QUEST Rr measurements of each quasar is shown in figure \ref{num_hist}.

Because some blazars (e.g. FSRQs) have quasar-like optical spectra, 
there may be blazars contaminating this spectroscopically identified 
quasar sample.  Blazar flux is dominated by beamed jet emission while 
quasar flux is thought to be primarily from the accretion disk;  before 
examining the objects in order to study the causes of quasar variability 
it is important to eliminate any known blazars from the list.  
Cross-checking with several blazar catalogs (\cite{1jansky}, 
\cite{bepposax}, \cite{cgrabs}, \cite{gamma1}, \cite{hewitt93}, 
\cite{massaro07}, \cite{roxa}, \cite{collinge05}, and \cite{vcv}) reveals 
47 blazars in the sample.  Furthermore, there are 92 objects that 
show dramatic variability in the QUEST Survey that is uncharacteristic 
of typical quasar behavior.  
They are seen to vary by more than 0.4 magnitudes over the course of the 
$\sim$3.5 years of our survey, which matches better with the 
behavior of known blazars than with quasars.  (The typical variability 
of blazars in the QUEST Survey will be studied in 
a future paper.)  Two of the 92 variables are among the 47 known 
blazars;  therefore we remove in all 137 objects from the list.

As mentioned earlier, extended objects will not be measured properly by the 
PSF photometry routines.  A conservative method of removing objects which 
are not pointlike in our data is to eliminate those that are found to be 
extended by SDSS.  SDSS goes roughly one magnitude deeper than QUEST 
in the r' filter and has seeing of 1.4'' as opposed to our 2.3''.  
The quasars in their photometric database are largely marked as 
morphologically pointlike past a redshift of 0.4, but mostly 
extended closer than that.  We therefore do not study those objects that 
have redshifts less than 0.4 so that we can be confident that the extent 
of the objects does not introduce significant systematic measurement errors.  
After this cut we are left with 22,825 quasars in our final sample.

Radio and X-ray information is known for small subsets of these quasars.  
We gather radio fluxes by matching the quasars with the FIRST Survey 
(\cite{becker95}).  A matching radius of 5'' yields 1,986 quasars with 
measured 1.4 GHz radio fluxes.  X-ray luminosities at 2 keV are taken from 
\cite{anderson07}, which carefully cross-matches SDSS quasars from their 
data release 5 with the ROSAT All-Sky Survey.  847 of the quasars in 
our sample have such X-ray measurements.

\section{The Structure Function}

There are many ways to parameterize variability.  The structure function 
($SF$) has been used by a number of studies to examine quasar variability 
(e.g. \cite{vandenberk04}, \cite{devries05}).  It is defined in several 
ways in the literature.  Two common definitions are as follows:

\begin{equation}SF^{(A)}(\tau) = \sqrt{<[m(t) - m(t-\tau)]^{2}> - <\sigma^{2}>}\label{sfa_eq}\end{equation}
\begin{equation}SF^{(B)}(\tau) = \sqrt{\frac{\pi}{2}<|m(t) - m(t-\tau)|>^{2} - <\sigma^{2}>}\label{sfb_eq}\end{equation}

$m(t) - m(t-\tau)$ is the difference in measured magnitudes of an object 
at two different times, where the times are separated by an interval $\tau$ 
in the quasar rest frame.  
$\sigma$ is the measurement error on the magnitude difference term 
so that the structure function measures only the intrinsic variations in the 
quasar flux.  $<X>$ denotes the mean value of $X$ measured for the set 
of objects.  The form given in equation \ref{sfb_eq} was introduced by 
\cite{diclemente96} because it is more robust to outliers than that given 
in equation \ref{sfa_eq}.  The two forms are equivalent if 
$[m(t)-m(t-\tau)]$ comes from a Gaussian distribution.  If this is true, 
then outliers in the data will tend to be measurement errors and equation 
\ref{sfb_eq} will accurately reflect the physics.  If the underlying 
variability distribution is not Gaussian, however, these two equations are 
not equivalent and equation \ref{sfa_eq} describes the variability statistics 
in the most straightforward way, as it is directly related to other 
statistical quantities such as the autocorrelation function and the variance.  
Because other studies have published results using variously $SF^{(A)}$ 
or $SF^{(B)}$, we have analyzed the QUEST data using both forms in order 
to compare with all of the available literature.  

A plot of the structure function vs. time lag has a characteristic shape based 
on the variability properties of the data.  For 
short timescales, where the intrinsic variability amplitude is smaller than the 
noise, the structure function will be flat with amplitude 
$\sqrt{2 \sigma^{2}}$.  For timescales shorter than those characteristic 
of the variability, but where variability can be measured, the structure 
function rises proportional to $\tau$.  At timescales typical of the variability 
the structure function has a shape that depends on the details of the 
system's lightcurves.  For timescales longer than those 
of the variability, the structure function is again flat, with amplitude 
$\sqrt{2 \sigma_{sys}^{2}}$ where $\sigma_{sys}$ is the average 
magnitude variation of the system.  These features would be evident in 
structure functions calculated from evenly sampled datasets over timescales 
much longer than those of the system's variability, in which 
edge effects and aliasing are negligible.  In finite, clumpy datasets 
such as ours the shape of the structure function may be affected by 
the data cadence, also called windowing.  This dependence is explored in appendix 
\ref{modelling_appendix};  in summary, windowing does not significantly alter the 
slope of the rising portion of the structure function, although it can 
introduce an artificial turnover at long time lags.

\section{Managing Correlated Quantities \label{binning_section}}

Quasar optical variability has been seen to depend on parameters other 
than timescale.  For example, the variability amplitude is well-known 
to be anticorrelated with optical luminosity (e.g. \cite{helfand01}; 
\cite{vandenberk04}).  It has also been seen that variability amplitude 
increases with quasar black hole mass (\cite{wold07}; \cite{wilhite08}).  Some 
correlations between quasar properties may simply be artifacts of the data, 
such as relationships between luminosity and redshift for objects discovered 
in flux-limited surveys.  In order to understand which intrinsic 
characteristics of the quasars most strongly influence the variability, 
we must treat the objects' parameters as independently as possible.  
We accomplish this using a method described below, which is similar to that 
used in \cite{vandenberk04} 

There are four basic quantities that we know for all of the quasars in our 
sample: rest frame time lag $\tau$ between measurements, optical luminosity 
$L$, estimated black hole mass $M_{BH}$, and redshift $z$.  There are known 
correlations between all 
of these parameters, either due to physical relationships or due to the 
biases of flux limited surveys, or perhaps a combination of both.  To 
avoid these complications and look at the dependence of variability on 
only black hole mass, for example, we would like to take a set of quasars 
with identical properties except for their black hole mass, and then 
examine how the variability differs between them.  To approximate this 
procedure, we have split each parameter's range into bins: six bins each 
in time lag, luminosity, mass, and redshift.  The choice of bin limits 
is shown in table \ref{sf_bins}.  Quasars with values outside the given 
ranges are not included in the analysis.  25\% of the quasar measurements 
are excluded due to the $L$, $M_{BH}$, and $z$ limits.  A further 35\% 
fall outside the $\tau$ bin limits, which are chosen to exclude data 
affected by systematics at low $\tau$ and windowing at high $\tau$.  The 
motivation for the exact $\tau$ limits are discussed in section 
\ref{sf_vs_time_section}.

\begin{table}
\begin{center}
\begin{tabular}{|l|c|c|c|c|}
\hline
& $\tau$ & $z$ & $M_{BH}$ & $L$ \\ \hline
Min & 100 & 0.4 & $10^{7}$ & 29 \\ \hline
& 150 & 0.7 & $5 \times 10^{8}$ & 29.75 \\ \cline{2-5}
Bin & 210 & 1.0 & $10^{9}$ & 30.25 \\ \cline{2-5}
Divisions & 280 & 1.3 & $2 \times 10^{9}$ & 30.75 \\ \cline{2-5}
& 350 & 1.6 & $3 \times 10^{9}$ & 31.15 \\ \cline{2-5}
& 425 & 1.9 & $5 \times 10^{9}$ & 31.55 \\ \hline
Max & 500 & 2.2 & $8 \times 10^{9}$ & 32 \\ \hline
\end{tabular}
\end{center}
\caption{Quasar sample bin limits introduced to handle correlated quantities orthogonally.  Units of $\tau$: days; $M_{BH}$: $\times M_{\odot}$; $L$: Log($\frac{\mathrm{erg}}{\mathrm{s} \cdot \mathrm{Hz}}$) measured at 2500 \AA.}
\label{sf_bins}
\end{table}

\begin{figure}
\begin{center}
\includegraphics[height=60mm]{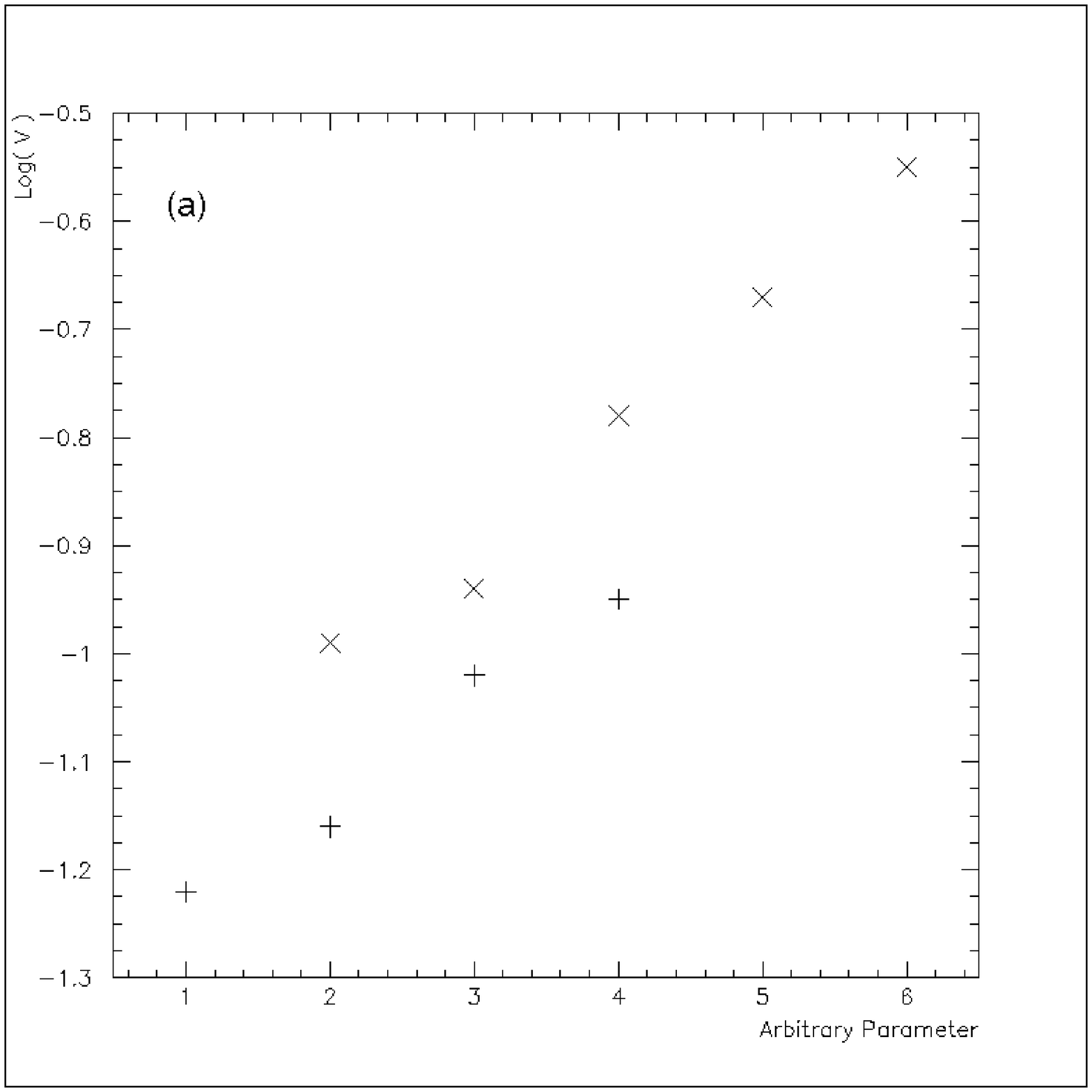}
\includegraphics[height=60mm]{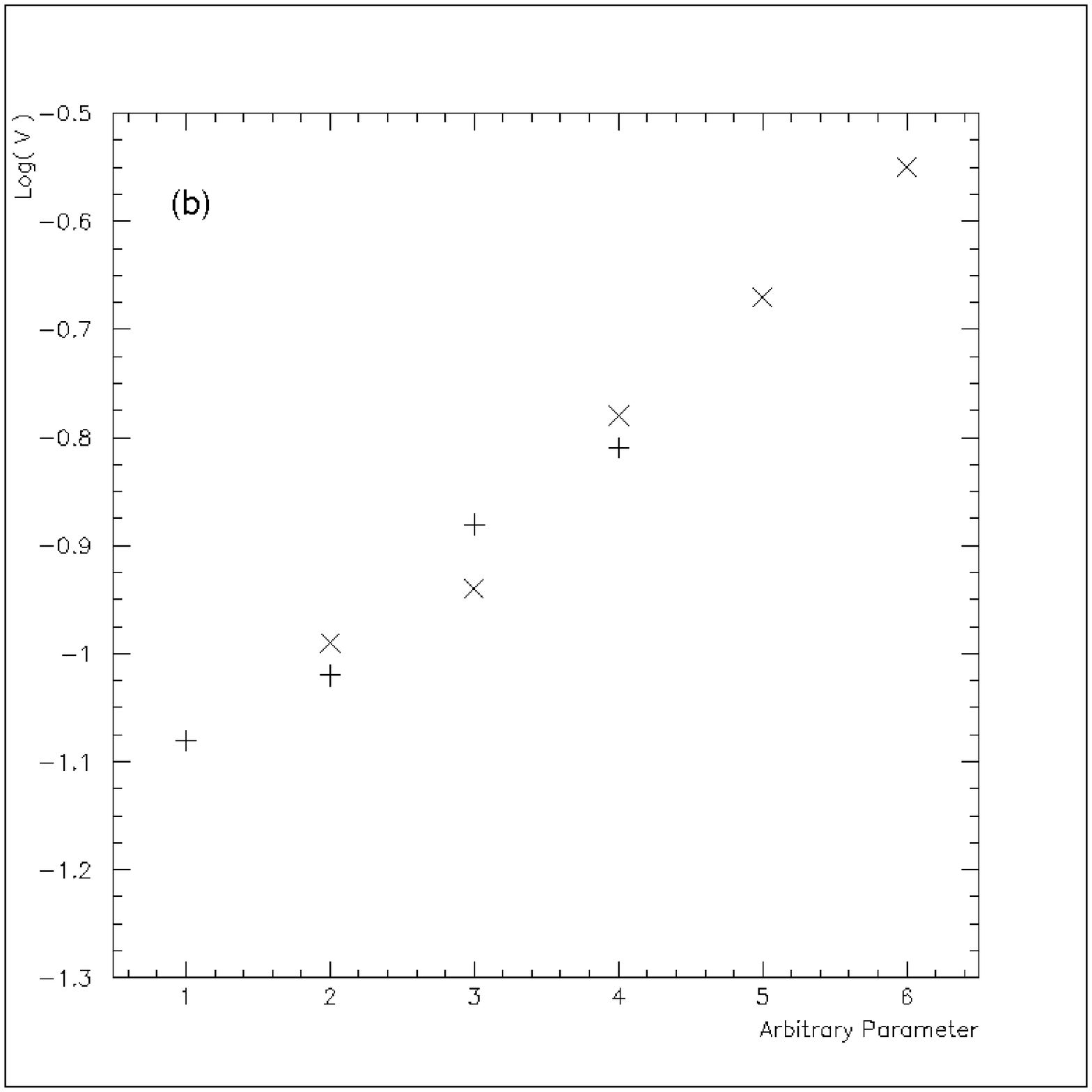}
\includegraphics[height=60mm]{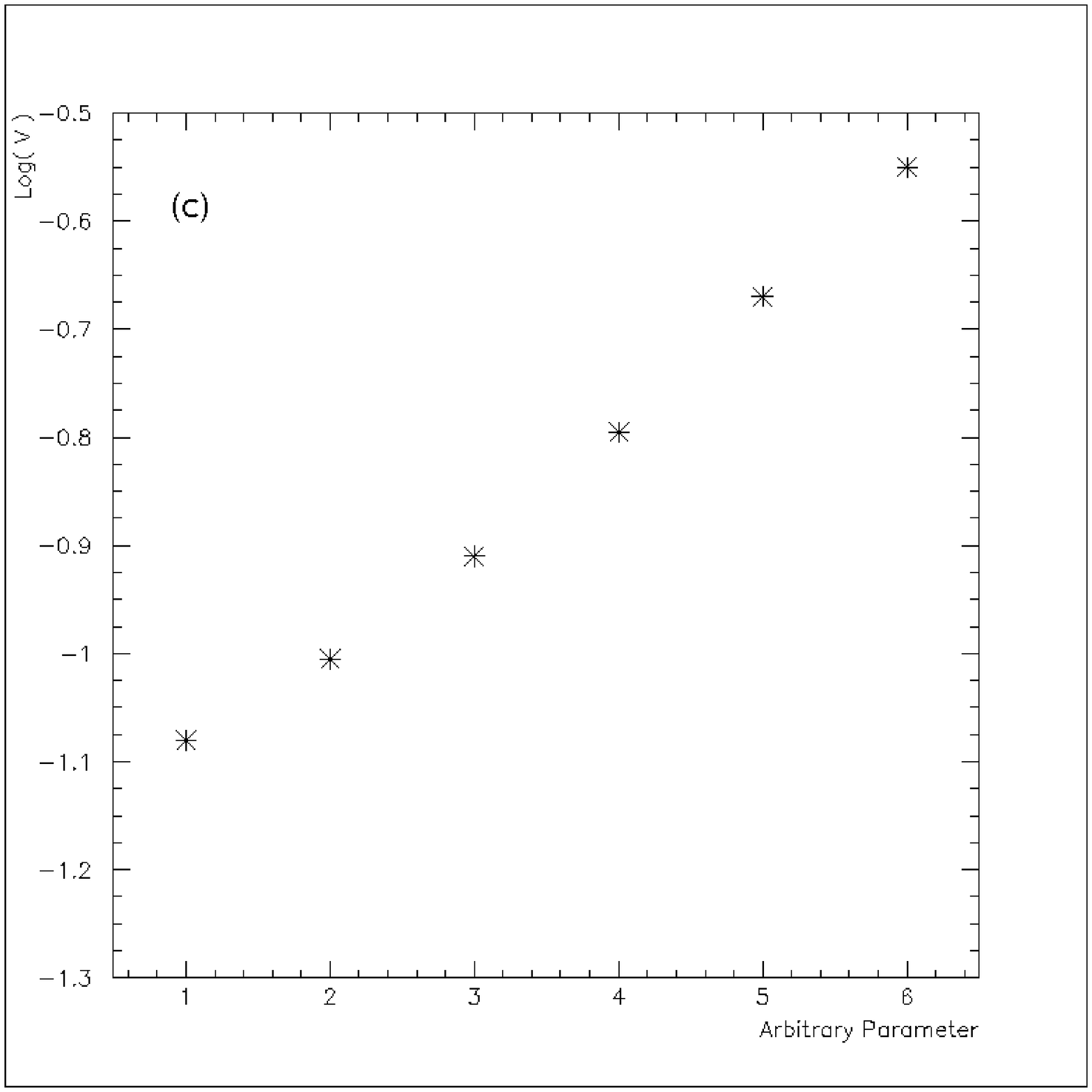}
\end{center}
\caption{Illustration of the normalization procedure.  $\times$ and + symbols represent two different data sets;  the + data are normalized to the $\times$ set in panel (b), and averaged to yield the * points in panel (c).}
\label{norm_example}
\end{figure}

In each multi-dimensional bin we calculate the average variability amplitude 
of the objects using the equation

\begin{equation} V = \sqrt{ <(\Delta m)^{2}> - <\sigma^{2}> } \label{v_eq}\end{equation}

which is identical to the structure function form $SF^{(A)}$ except it is 
not a function of $\tau$ but instead compares all available measurements 
of the same quasar.  Then, holding constant the bin indices for time lag, 
luminosity, and redshift, we can legitimately compare 
the six bins of black hole mass.  This procedure yields $6 \times 6 \times 
6 = 216$ possible six-point plots of variability  amplitude $V$ vs. mass, or 1296 
possible $V$ values.  Most of the combinations of parameters do 
not describe many objects in our sample.  In order to determine a 
statistically significant value of $V$ , we require 100 measurement pairs 
in a bin before including it in the analysis.  The number of measured $V$s 
is therefore much smaller than 1296 due to statistics.  In fact, we measure 
148 well-sampled $V$ values.   To examine the overall behavior of the 
variability amplitude in terms of the mass we can normalize 
the six-point measured trends together and 
average the resulting normalized data in each mass bin in order to find a 
simple, meaningful result of how the variability scales with the quasar's 
mass (or any of the other parameters).  
Because the data are arbitrarily normalized to the set with the best 
statistics, the exact values for $V$ should not be taken seriously.  The 
shape of the trend is the important result.  The normalization process 
is illustrated in figure \ref{norm_example}.  Panel (a) shows two 
unnormalized datasets of log($V$) versus an example parameter.  For simplicity, 
each point is assumed to 
have equal uncertainties, which are not shown.  The $+$ dataset will be 
normalized to the $\times$ dataset, as the $\times$ dataset has the higher 
statistics.  The normalization consists 
of an additive constant on the log scale, in order to preserve the power law 
shapes observed in the $V$ vs. time and luminosity trends.  The constant is 
determined by minimizing the chi squared difference between the values from 
the two datasets in the same bin, for the bins where there exist data from both 
sets.  Panel (b) in figure \ref{norm_example} 
shows the data after the $+$ data are shifted by one constant such that they 
best agree with the $\times$ data.  
Panel (c) shows the final result after averaging the 
normalized data in each bin.

\section{Results}

\subsection{Variability vs. Optical Luminosity \label{luminosity_section}}

\begin{figure}
\begin{center}
\plotone{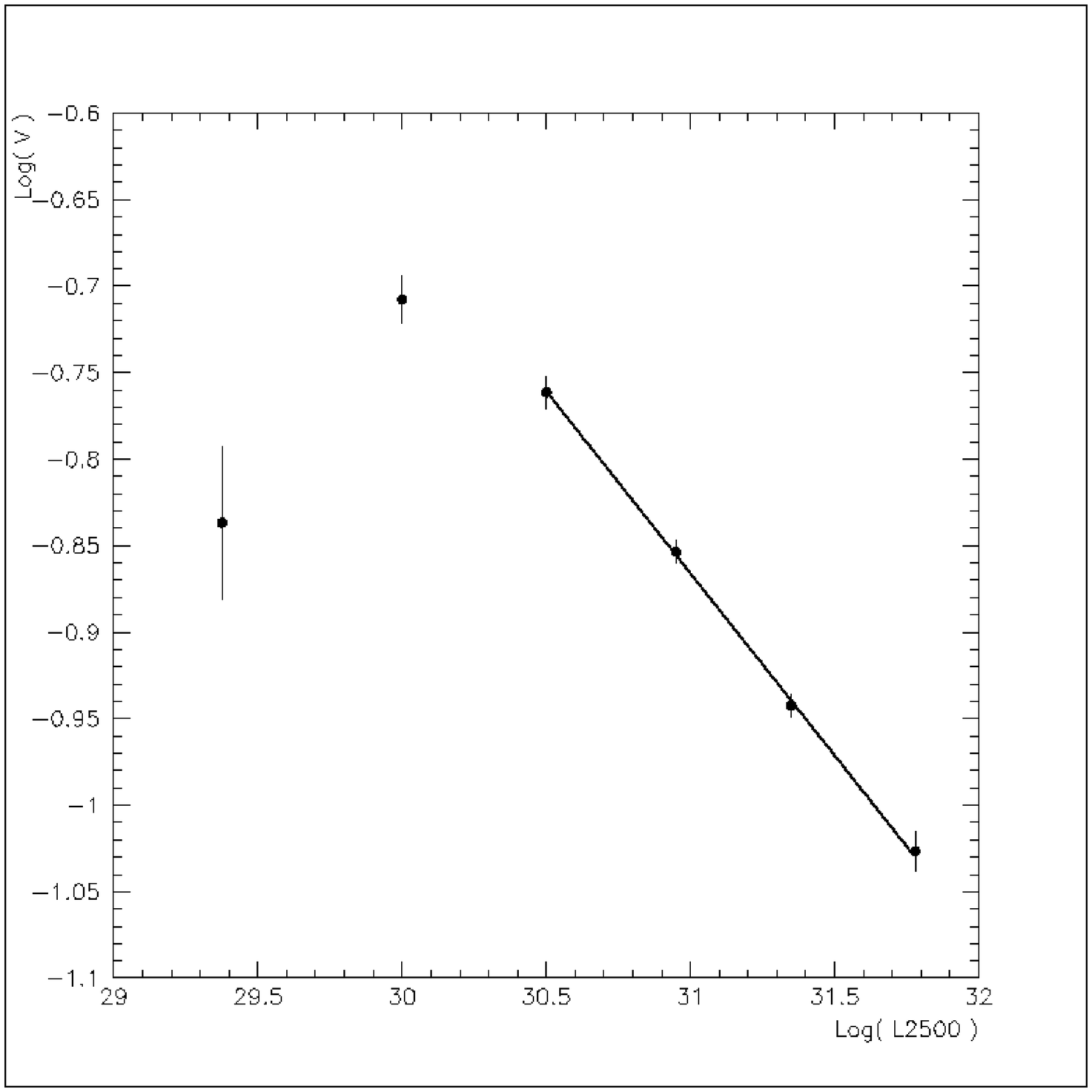}
\caption{Logarithm of quasar variability $V$ vs. logarithm of optical luminosity at 2500 \AA, with a linear fit of the brightest four points.}
\label{sf_luminosity}
\end{center}
\end{figure}

Log($V$) vs. log($L$) for the quasar sample, where $L$ is the 
luminosity of the quasar calculated at a rest frame wavelength of 2500 \AA, 
is shown in figure \ref{sf_luminosity}.  2500 \AA\ is chosen because it is a 
wavelength dominated by continuum emission that, for the redshift range of 
the sample, lies predominantly in the high-throughput SDSS g, r, or i 
filters.  The luminosity at rest frame 2500 \AA\ is determined by convolving 
a redshifted composite quasar spectrum taken from \cite{compositespectrum} 
with the SDSS filter curves and using the SDSS broadband flux measurements to 
normalize the composite spectrum's amplitude.  
The flux at any particular wavelength is then given by the 
normalized composite spectrum, and can be converted to luminosity using the 
object's redshift and the cosmological parameters which we assume to be 
$\Omega = 1, \Omega_{\Lambda} = 0.7, \Omega_{M} = 0.3, H_{0} = 71 
\frac{\mathrm{km}}{\mathrm{s} \cdot \mathrm{Mpc}}$.  
The downturn in figure \ref{sf_luminosity} at low luminosities 
is most likely an artifact of our magnitude limit;  if a faint quasar 
becomes dimmer we may not see it.  We will then not observe the whole range 
of variability for the faintest objects.  There are relatively few statistics 
in the first two luminosity bins, contributing to a large systematic error 
as well:  the first bins contain 4 and 10 measurements, respectively, 
compared to between 31 and 63 in the remaining bins.  
We fit a line to the four brightest points in 
figure \ref{sf_luminosity} and find a slope of $0.205 \pm 0.002$.
The reduced $\chi^{2}$ of the fit is very small: $\chi^{2} = 0.1735$.  

The errors on the points in this and all subsequent plots are the 1$\sigma$ 
standard deviation of values obtained by analyzing subsets of the total 
sample.  

The observed trend of average variability amplitude with luminosity implies 
that we do not fully measure the variability of low-luminosity objects.  For this 
reason, quasars with luminosity less than $10^{30.5} \frac{\mathrm{erg}}{\mathrm{s} \cdot \mathrm{Hz}}$ 
are not included in the remaining analysis unless stated otherwise.

\subsection{Quasar Structure Function vs. Time \label{sf_vs_time_section}}

\begin{figure}
\begin{center}
\plottwo{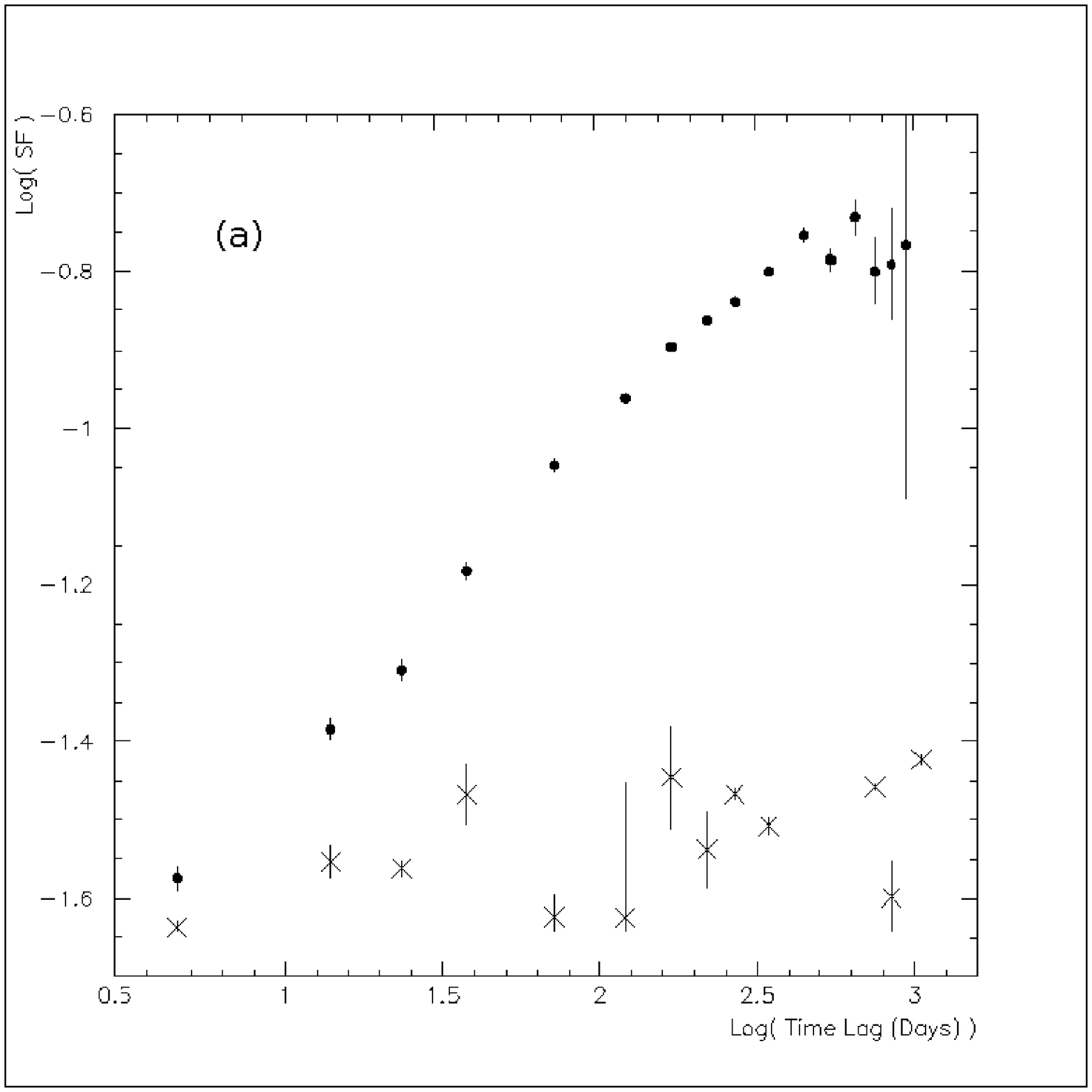}{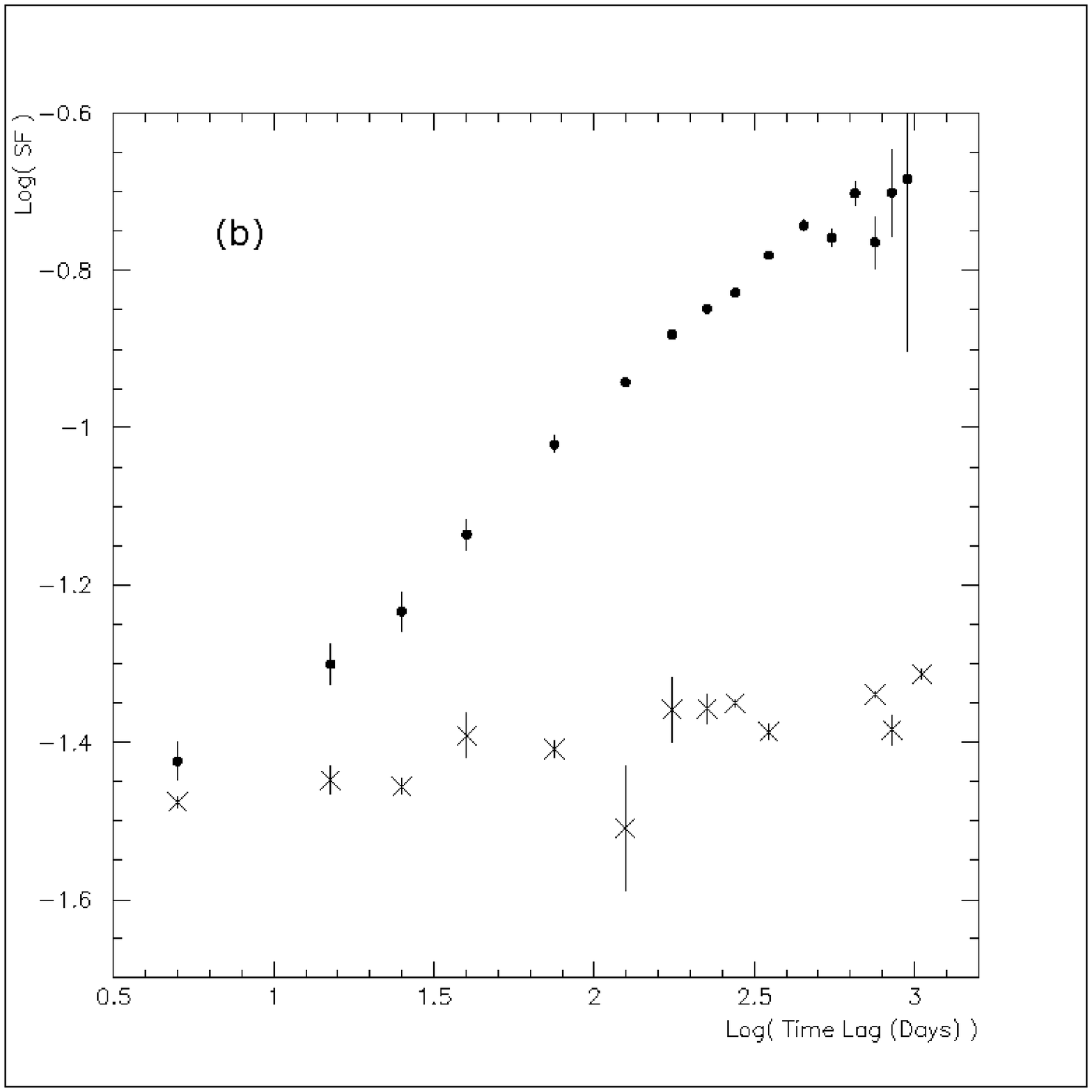}
\caption{QUEST logarithmic structure functions vs. logarithmic rest frame time lag.  (a): $SF^{(A)}$.  (b): $SF^{(B)}$.  Solid points: quasar results. $\times$: sample pointlike objects.}
\label{sf_time_results_fig}
\end{center}
\end{figure}

In order to show results comparable to other published work we have 
calculated the 
structure function, binning only in rest frame time lag and not in other 
parameters such as luminosity, using both forms $SF^{(A)}$ and $SF^{(B)}$.  
This analysis does include quasars of all luminosities.  
The results are shown as solid points in figure \ref{sf_time_results_fig}, 
along with, as $\times$s, analogous results for a random sample of 
pointlike objects.  The quasar structure function rises approximately 
as a power law.  The turnover we see at long 
time lags is due not to intrinsic qualities of the variability, but to 
windowing effects as described in appendix \ref{modelling_appendix}.  
The power law index, often called the structure 
function slope since the data tend to be plotted on a log-log scale, 
is measured to be 0.357 $\pm$ 0.014 using $SF^{(A)}$, and 
0.3607 $\pm$ 0.0075 using $SF^{(B)}$.  
The two results agree with each other to within $1\sigma$;  
we adopt the form $SF^{(A)}$ for use in the rest of the structure function 
analyses.

In the case of $SF^{(A)}$ we iteratively clip the data to reduce the 
effects of measurement errors and atypical objects.  
To ensure that the structure funtion slope is not sensitive to the details of 
the clipping, we only fit the points in $SF^{(A)}$ whose clipping quickly converges. 
The time lag bins with low variability amplitude are most affected by outliers;  
therefore only longer time lag bins, with $\tau > 100$ days, are included in 
the fit.  The presence of low-$\tau$ outliers may partially be due to color 
terms in the data, which contribute at a level of $SF \sim 0.05$.  The last 
several structure function points suffer from edge effects due to the QUEST 
data cadence, as examined in 
appendix \ref{modelling_appendix}.  So, $SF^{(A)}$ is fit 
between rest frame time lags of 100 and 500 days;  these limits are also used 
in the binned $V(\tau)$ analysis described in section \ref{binning_section}.  
The $SF^{(B)}$ fit 
includes data at smaller time lags since this form is designed to minimize the 
effects of outliers in the data.  For the $SF^{(B)}$ result, then, we fit 
all time lags shorter than 600 days, after which the windowing effects 
become prominent.

The structure functions of a random sample of pointlike objects, also shown in 
figure \ref{sf_time_results_fig}, are flat.  
The errors on these data vary substantially from point to point because the 
objects are assumed to be stars at zero redshift, so there are low 
statistics at multiples of half-year rest frame time lags.  

We have also calculated the quasar structure function vs. rest frame 
time lag using the multi-dimensional binning and normalizing technique 
described in section \ref{binning_section}.  This procedure, when 
performed using quasars of all luminosities, yields a 
structure function slope of 0.392 $\pm$ 0.022, which agrees with our 
other measurements to within the 1$\sigma$ for the $SF^{(A)}$ result 
and 1.1$\sigma$ for the $SF^{(B)}$ result.  We expect this 
similarity because the correlation between the rest frame time lag between 
observations of a quasar and the quasar's 
redshift is most pronounced at the extremes of the time lag distribution 
and we are only fitting a middle range.  Furthermore, there is no reason 
why the time lag between our observations of certain quasars would 
depend on intrinsic properties like their mass or luminosity.  Therefore 
we expect the fit to be minimally affected by the binning.  
When restricting the analysis to only those quasars with luminosity at 
rest frame 2500 \AA\ greater than $10^{30.5} \frac{\mathrm{erg}}{\mathrm{s} 
\cdot \mathrm{Hz}}$, as motivated in section \ref{luminosity_section}, 
the measured slope increases to 0.432 $\pm$ 0.024.  This latter 
result is a better measurement of the quasars' variability behavior 
with time lag, as it excludes objects whose faintest states appear to be 
below our flux limits.

\subsubsection{Lightcurve Asymmetry}

\begin{figure}
\begin{center}
\plotone{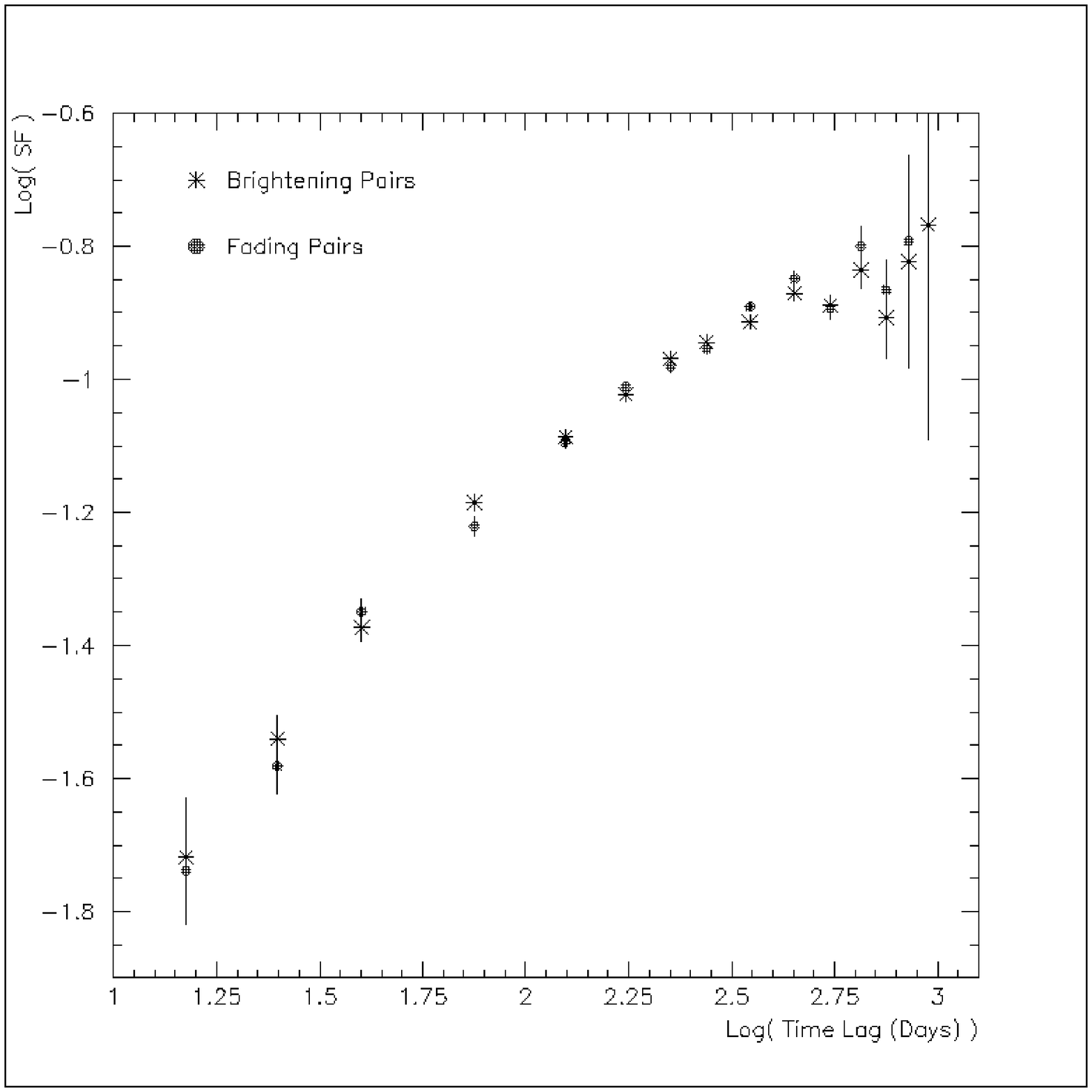}
\caption{Logarithmic quasar structure function vs. logarithmic rest frame time lag using only brightening (asterisks) and fading (solid circles) magnitude pairs.}
\label{sf_time_asymmetry}
\end{center}
\end{figure}

The shape of quasar flares holds information about the flares' physical source.  
Because there are typically only four QUEST measurements of each quasar, 
we can not study the detailed lightcurve shapes of the 23,000 quasars 
individually.  However, we can examine any flare asymmetry by calculating 
the structure functions from brightening and fading subsets of the data, 
i.e. the measurement pairs for which $m(t)>m(t+\tau)$ and 
$m(t)<m(t+\tau)$, respectively.  
If the variability were due to flares which brightened over a month and 
faded over a year, a structure function calculated using only brightening 
magnitude pairs would have more power on a month timescale and less on a 
year timescale than one calculated using only fading magnitude pairs.  
Figure \ref{sf_time_asymmetry} shows the brightening and fading structure 
functions, 
calculated without multi-dimensional binning, for all quasar luminosities, 
using $SF^{(A)}$.  
The asterisks illustrate the brightening measurements and the solid 
points reflect the fading ones.  There is no evidence for any flare 
asymmetries over the timescales that we measure;  the two structure 
functions are consistent with each other to within errors.

\subsection{Variability vs. Black Hole Mass}

\begin{figure}
\begin{center}
\plotone{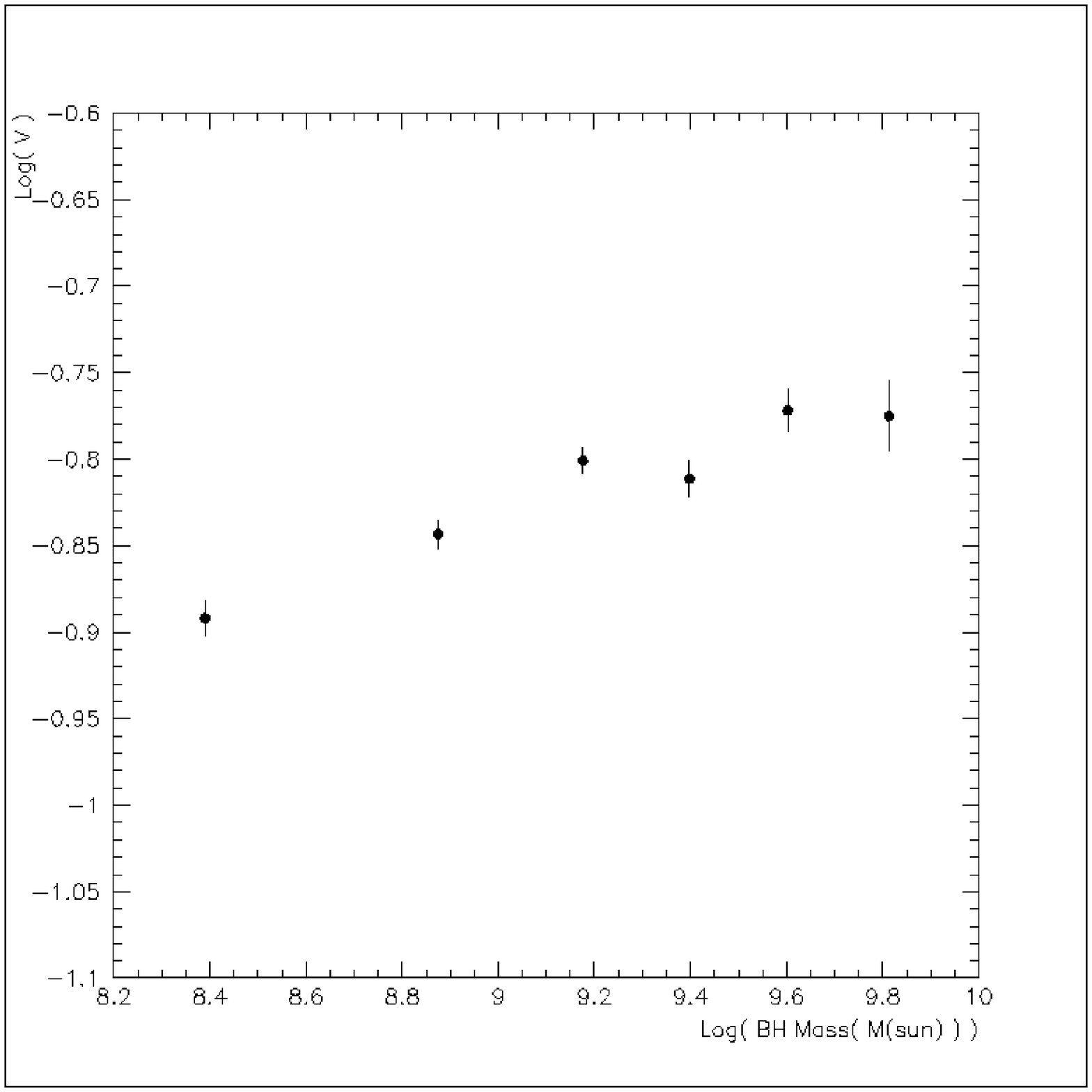}
\caption{Logarithm of quasar variability $V$ vs. estimated black hole mass.}
\label{sf_bhmass}
\end{center}
\end{figure}

Given a quasar spectrum, one can estimate the black hole mass.  The method 
we use is based on that of \cite{kaspi00}, further discussed in 
e.g. \cite{shields03} and \cite{salviander07}, 
and thought to be good to a factor of 3.  Given that our quasar sample spans 
two orders of magnitude in mass, the error associated with this calculation 
will not mask a global trend.  
We calculate each quasar's black hole mass using the equation

\begin{eqnarray}
M & = & (10^{7.69} M_{\odot} ) \nonumber\\ & \times & \frac{ [\sigma_{\mathrm{H\beta}}, \sigma_{\mathrm{MgII}}] }  {3000 \: \mathrm{km s^{-1}} } \\ & \times & \sqrt{ \frac{ \lambda L_{\lambda 5100} }{ (1044 \: \mathrm{erg s^{-1}}) } } .\nonumber
\end{eqnarray}

where [$\sigma_{\mathrm{H\beta}}, \sigma_{\mathrm{MgII}}]$ is the full width at 
half maximum of the $\mathrm{H_{\beta}}$ or $\mathrm{MgII}$ emission line, 
depending on which is available in the SDSS database.  If both are measured, 
we use the average of the two values.  
$L_{\lambda 5100}$, the luminosity of a quasar at 5100 \AA, is calculated 
from the SDSS broadband fluxes in the same manner as $L_{\lambda 2500}$, 
as described in section \ref{luminosity_section}.  
The logarithm of the quasar variability amplitude $V$ vs. the logarithm of 
the estimated black hole mass, calculated after multi-dimensional 
binning, normalization, and averaging, is shown in figure \ref{sf_bhmass}.  
We see a significant increase in quasar variability amplitude with black 
hole mass over the entire mass range studied.  
However, it is a much smaller effect than that seen with respect to time 
lag or luminosity.

It is important to note that the relationship between variability amplitude 
and luminosity entirely masks the trend with black hole mass if the results 
are calculated without regard to the correlation between the quasars' 
properties.  The brightest quasars are usually the most massive, which 
leads to the apparent result that the massive quasars vary least.  Once 
the luminosity effects are separated out, however, the trend completely 
reverses to show an increase in variability amplitude with mass.

Because physical timescales such as the accretion, dynamical, and thermal 
timescales of AGN disks scale with black hole mass, it is plausible that 
optical variability timescales might differ between AGN of different masses.  
However, the average variability amplitude $V$ versus time, calculated 
separately using low, medium, and high mass subsets of the data, yield 
consistent results.

\subsection{Variability vs. Eddington Ratio}

The Eddington ratio is the object's bolometric luminosity 
$L_{\mathrm{bol}}$ divided by its Eddington luminosity $L_{\mathrm{edd}}$, 
and is a useful parameter describing the accretion rate of the central engine.  
It can be estimated using the equation

\begin{equation}
E \equiv \frac{L_{\mathrm{bol}}}{L_{\mathrm{edd}}} \approx  \frac{9 \times \lambda L_{\lambda 5100}}{1.3 \times 10^{38} (M_{BH}/M_{\odot})}
\end{equation}

where the approximation for the bolometric luminosity is that made 
by \cite{kaspi00}.

Because the black hole mass estimate, $M_{BH}$, is uncertain to roughly a 
factor of three, the Eddington ratio is as well.  $M_{BH}$ is thought to 
be an overestimate 
as often as an underestimate, so the net effect on the 
results is a smearing of the values rather than a systematic shift.  
The consequence of any error in $M_{BH}$ will therefore simply 
cause a flattening of the results rather than introduce a spurious trend.  

To calculate the dependence of variability amplitude on Eddington ratio we 
reorganize the data, substituting both the black hole mass and luminosity binning 
for one binning in Eddington ratio.  The dependencies on time lag and 
redshift continue to be normalized out as before.

\begin{figure}
\begin{center}
\plotone{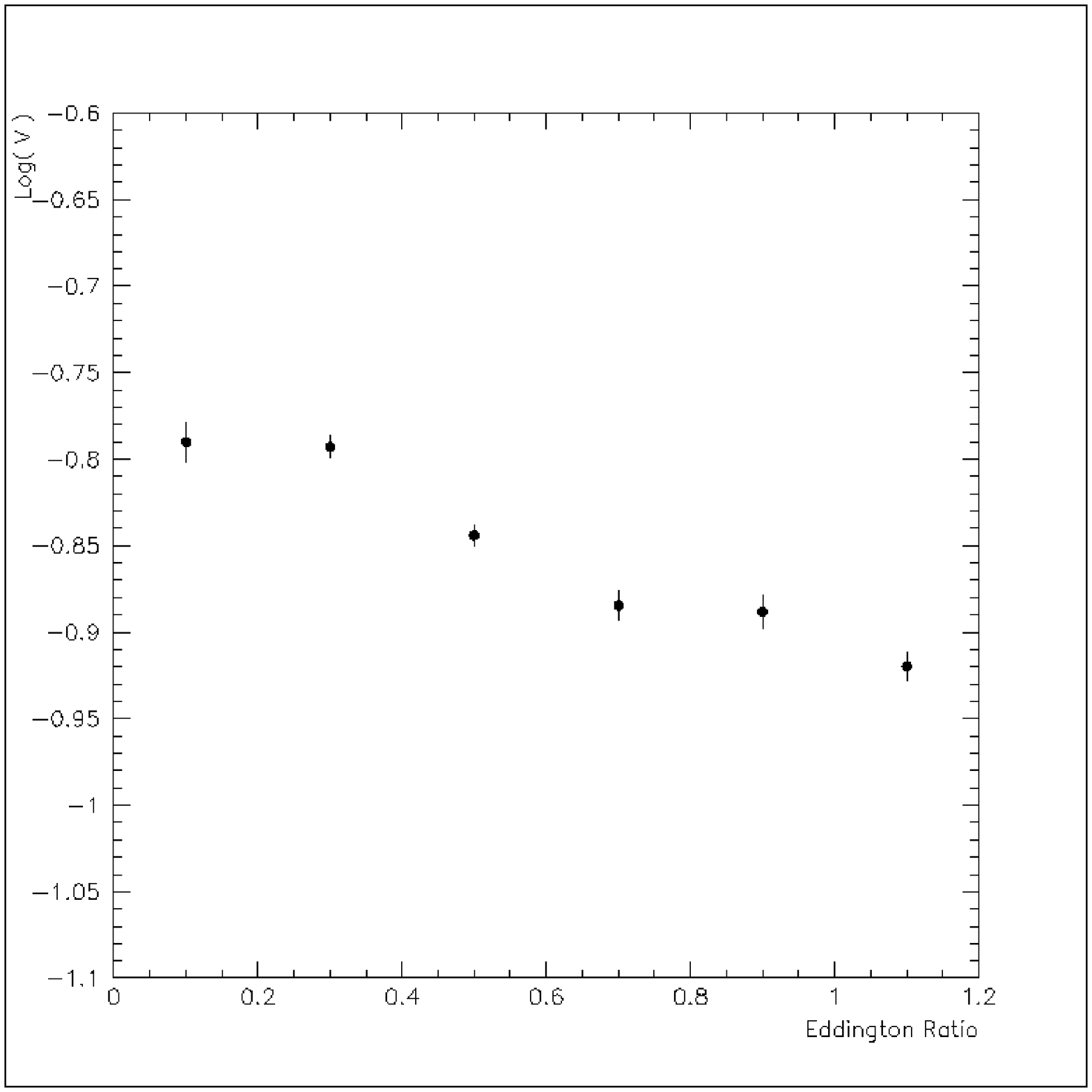}
\caption{Logarithm of quasar variability $V$ vs. Eddington ratio.}
\label{sf_er}
\end{center}
\end{figure}

The resulting plot of Log($V$) versus Eddington ratio is shown in 
figure \ref{sf_er}.  The average variability amplitude decreases with 
increasing ratio.  The calculated 
Eddington ratios of the quasars are predominantly between zero and 
one, but include a very long, small tail out to large values, which 
may be due to poor luminosity or black hole mass measurements.  
The point shown at 1.1 on the $x$ axis is in fact calculated using all 
quasars with estimated Eddington ratios between one and ten.  

\begin{figure}
\begin{center}
\plotone{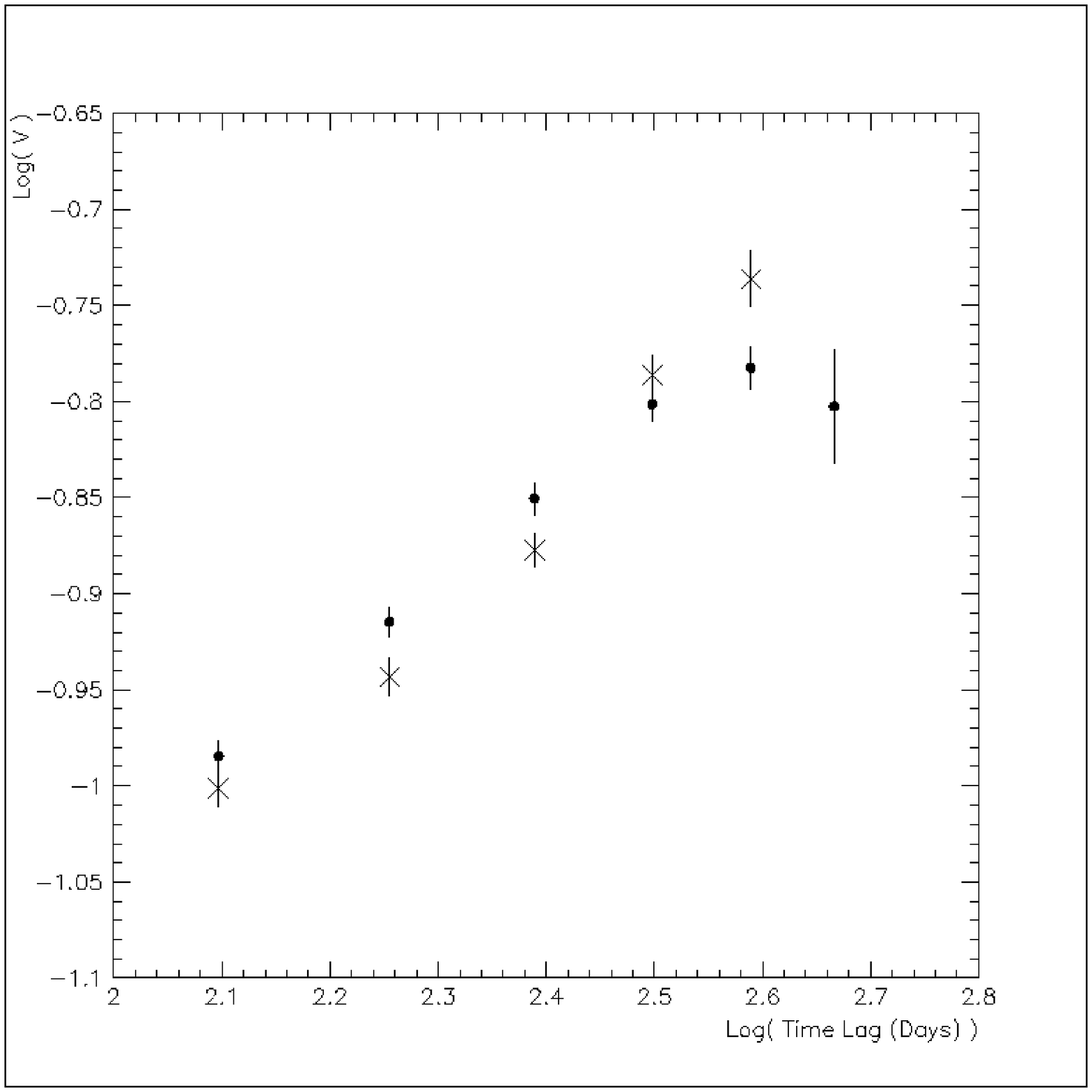}
\caption{Logarithm of quasar variability vs. logarithm of rest frame time lag for only low Eddington ratio (solid cicles) and high Eddington ratio ($\times$) objects.}
\label{sf_time_byer}
\end{center}
\end{figure}

To examine the timescales on which the quasars with different Eddington 
ratios vary, we have calculated the structure function versus time 
for only objects with low Eddington ratios between 0 and 0.4, and again 
for only objects with high Eddington ratios between 0.6 and 1.  
The results are shown in figure \ref{sf_time_byer}, where the solid 
circles show the data with low Eddington ratios and the $\times$ symbols 
indicate data with high Eddington ratios.  
At time lags shorter than roughly 300 days, 
the low Eddington ratio quasars are consistently more variable than those 
with high ratios.  The trend is reversed at longer time lags.

\subsection{Variability vs. X-ray Emission}

Because we have X-ray information for only a small subset of the quasars, 
we split the X-ray values into only three bins: bin 0 includes all objects 
with no X-ray information, bin 1 holds objects with X-ray slope 
$\alpha_{ox}$ greater than 1.35, and bin 2 holds those with $\alpha_{ox}$ 
less than 1.35.  $\alpha_{ox}$ is defined as the exponent of a hypothetical 
power law interpolated between the 2 keV X-ray and the 2500 \AA\ optical 
measurements of the object, as is used in works such as \cite{anderson07}.  
Bin 2 is X-ray louder than bin 1, with the 
division chosen so that the two bins have roughly equal numbers of objects.  
It is important to note that since we are normalizing out the effects of 
optical luminosity, our X-ray results are closer to a variability 
trend with X-ray luminosity rather than with X-ray loudness.  

Splitting the X-ray detected quasars into subsets by $\alpha_{ox}$, time 
lag, redshift, luminosity, and black hole mass leaves very few objects in 
each bin.  To gain enough statistics to calculate $V$, we collapse the 
time lag and redshift bins and only divide the quasars by $\alpha_{ox}$, 
luminosity and mass.  Furthermore, we insist on only 50 measurement pairs 
(rather than 100) as a minimum for calculating $V$.  If the X-ray loudness 
is correlated with redshift (or, improbably, with time lag between QUEST 
observations) then the relation we measure between variability and 
$\alpha_{ox}$ will be confused with any relation between variability and 
redshift or time lag.  The redshift distributions of the X-ray louder and 
X-ray quieter quasars are quite similar, however, so any variability trends 
with redshift will contribute similarly to each bin.  The correlations 
between X-ray loudness and optical luminosity and mass are indeed normalized 
out of the results, which are given in table \ref{sf_x_table}.

\begin{table}
\begin{center}
\begin{tabular}{|l|c|}
\hline
X-ray Emission & Log($V$) \\\hline
Unknown & -0.941 $\pm$ 0.006 \\
Fainter Subset & -0.813 $\pm$ 0.024 \\
Brighter Subset & -1.022 $\pm$ 0.049 \\\hline
\end{tabular}
\end{center}
\caption{Variability Log($V$) for quasars of different X-ray emission.  
Bins correspond roughly to X-ray luminosity, as described in the text.}
\label{sf_x_table}
\end{table}

These results show a 2.9$\sigma$ anticorrelation between optical variability 
amplitude and X-ray emission.  The quasars for which we have no X-ray information 
have an average optical variability amplitude intermediate between the values for 
the X-ray fainter and X-ray brighter samples.

\subsection{Variability vs. Radio Emission}

As is the case with the X-ray data, only a small subset of the quasars 
has radio measurements.  To calculate $V$ with the available statistics we 
make the same concessions as in the X-ray analysis:  computing $V$ with 
a minimum of 50 measurement pairs and eliminating the time lag and redshift 
binning.  Any correlation between redshift and radio loudness will not be 
removed; therefore if the redshift and radio properties are correlated then 
the variability dependence on these parameters will be confused.  However, 
the redshift distribution of the radio subsamples are similar, 
implying no strong relationship between radio loudness and redshift for 
our sample.  The quasars are divided into three radio bins: bin 0 includes 
all objects with no radio information, bin 1 has objects with ratios $R$ of 
5 GHz radio to 4500 \AA\ optical luminosity below 20, and bin 2 has objects 
with $R$ above 20.  The definition of radio loudness $R$ 
follows that of authors such as \cite{peterson}, and the break point 
between the two radio subsets was chosen so that each bin has roughly 
equal numbers of objects.  
As with the X-ray sample, it should be noted that once the 
normalization is performed, the measurements are divided according to 
a measure closer to luminosity than to loudness.  

The logarithm of the quasar variability amplitude $V$ for samples with 
different radio emission is given in table \ref{sf_r_table}.  
The three measurements agree within their errors;  we therefore 
see no evidence for a change in the average optical variability 
amplitude with increasing radio luminosity.  

\begin{table}
\begin{center}
\begin{tabular}{|l|c|}
\hline
Radio Emission & Log($V$) \\\hline
Unknown & -0.870 $\pm$ 0.005 \\
Fainter Subset & -0.880 $\pm$ 0.027 \\
Brighter Subset & -0.880 $\pm$ 0.028 \\\hline
\end{tabular}
\end{center}
\caption{Variability Log($V$) for quasars with different radio emission.  
Bins correspond roughly to radio luminosity, as described in the text.}
\label{sf_r_table}
\end{table}

\begin{figure}
\begin{center}
\plottwo{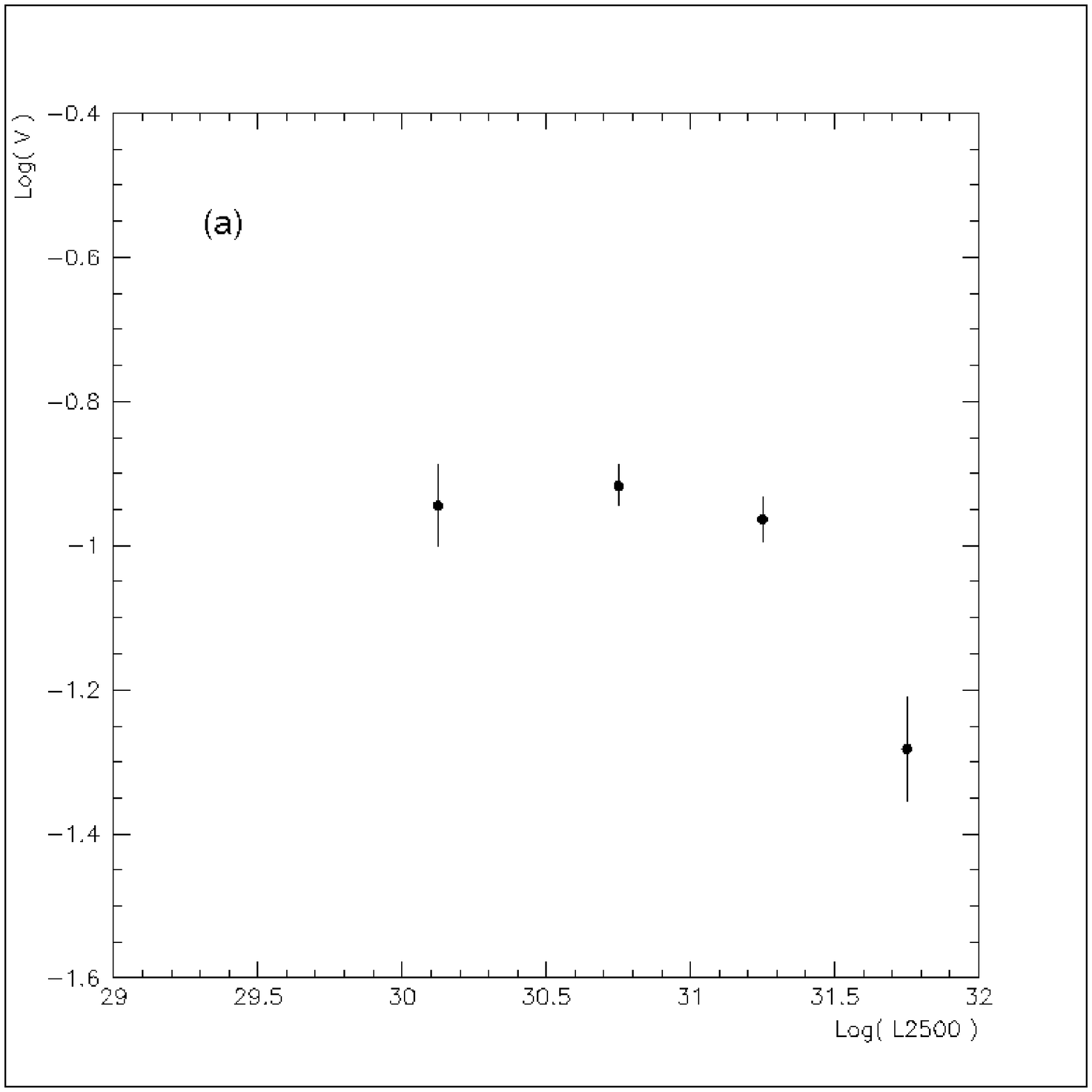}{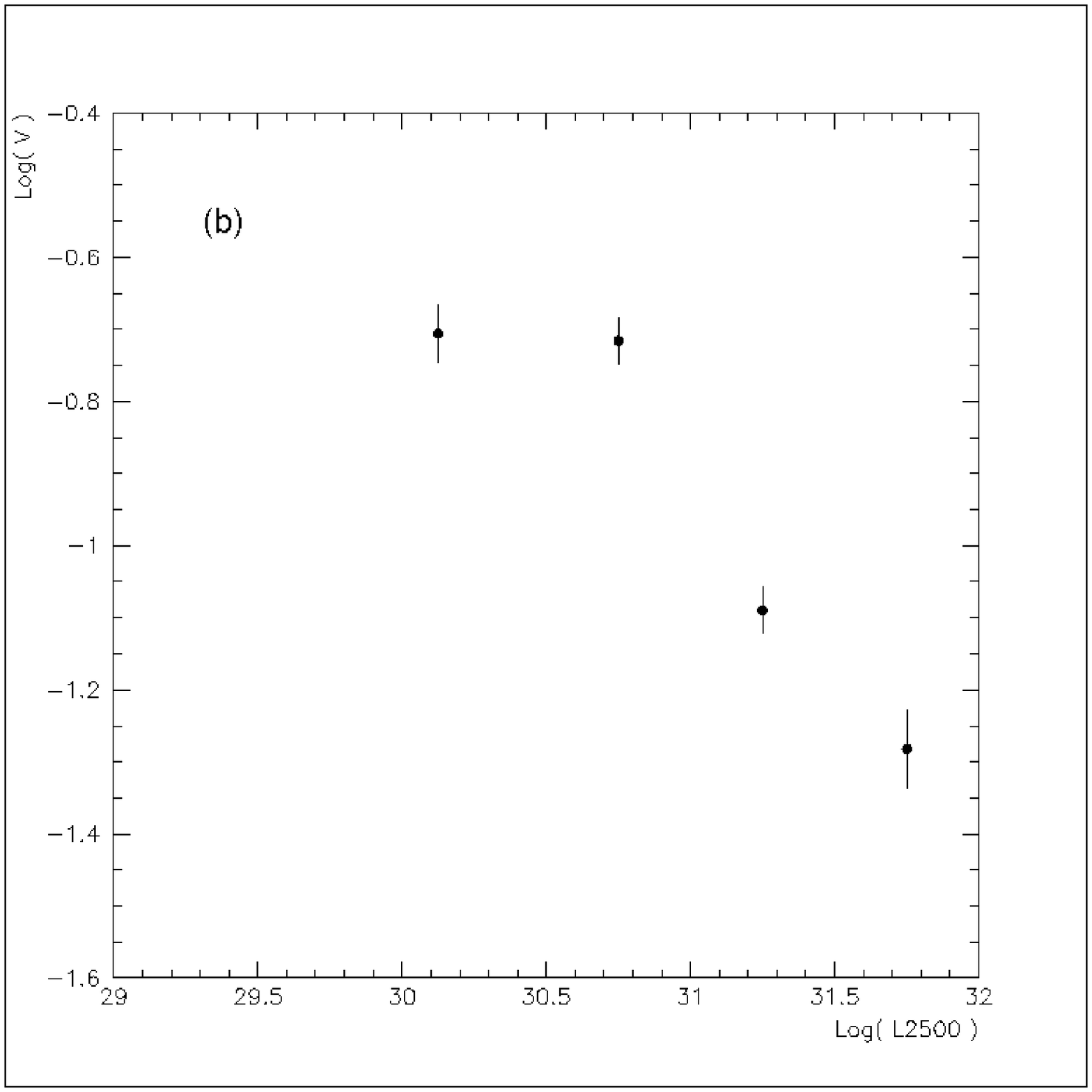}
\caption{Logarithm of quasar variability $V$ vs. logarithm of optical luminosity at 2500 \AA for objects with radio measurements.  (a): objects with radio to optical ratio less than 20.  (b): objects with ratios above 20.}
\label{sf_l2500_radio}
\end{center}
\end{figure}

The observed trend of decreasing variability amplitude with increasing optical 
luminosity has previously been seen only in radio quiet quasars, 
prompting \cite{helfand01} to hypothesize different variability mechanisms 
for radio loud and quiet objects.  To see if the luminosity trend that we 
observe is the same for radio louder and quieter quasars in our sample we 
examine the log($V$) vs. optical luminosity using 
only objects in one radio bin at a time.  The results are shown in figure 
\ref{sf_l2500_radio}:  the first panel includes radio quieter quasars 
(those with $R<20$); the second panel includes 
radio louder ones (with $R>20$).  We see that the 
trend indeed exists in both samples, even appearing stronger in the 
radio loud objects.  

\section{Discussion}

\subsection{Structure Function vs. Time Lag}

The slope of the structure function versus rest frame time lag has been 
measured for several large ensembles of quasars, using different types 
of data, different time scales, and different definitions of the structure 
function.  It is therefore not straightforward to compare all of the 
results.  

We have calculated the structure function slope in four ways: first, 
using equation \ref{sfa_eq} and binning the data only in time lag; 
second, using equation \ref{sfb_eq} and binning the data only in time 
lag; third, using equation \ref{sfa_eq} and normalizing the data after 
binning it in luminosity, mass, and redshift; and fourth, identical to 
the third method but only including quasars with 2500 \AA\ luminosity 
greater than $10^{30.5} \frac{\mathrm{erg}}{\mathrm{s} \cdot \mathrm{Hz}}$, 
at which level we are confident we can measure 
the full range of typical variability.  The last result is the most 
robust measurement, but if faint quasars have different variability 
timescales than bright ones then we are only observing one end of the 
behavior.

Our results are compared to others from the literature in table 
\ref{results_comparison_table}.  
Because our results using structure functions defined by equations 
\ref{sfa_eq} and \ref{sfb_eq} are consistent with each other, we do 
not distinguish in the table between results using the different forms.  
In table \ref{results_comparison_table}, QUEST2 refers to this work.  
QUEST1 refers to \cite{adam}, who studied 
933 quasars using the QUEST I Variability Survey.  Each quasar had roughly 
25 measurements in the Johnson R band, and the structure function slope 
was fit between 50 and 600 days.  SDSS + Plates 
refers to work by \cite{devries05}, in which SDSS data were 
compared with POSS and GSC2 archival data in order to generate a dataset 
that spanned roughly 50 years.  The structure function slope was fit 
over timescales of one to twenty years, over which the data appears 
to follow a power law.  
After these timescales their structure function 
appears to flatten, although the authors note that the longest timescale 
data is noisy and does not indicate a significant plateau in the structure 
function.  However, \cite{sesar06} also studied SDSS and POSS data that 
spanned several decades and saw that the variability amplitude on decade 
timescales is smaller than that expected from an extrapolation of power law 
results fit using $\sim$3 years of data.  
SDSS + Spectra in 
table \ref{results_comparison_table} refers to \cite{vandenberk04}, 
who compared SDSS broadband measurements to spectrophotometry 
calculated from the SDSS spectra which identified the objects as 
quasars.  They calculated the structure function slope before and 
after binning and normalizing their data in a manner similar to 
our method.  SDSS Equ. denotes the \cite{wilhite08} study of quasars 
with $\sim$10 SDSS measurements in a 278 square degree equatorial region.  
The study focuses mainly on the links between variability and mass, 
luminosity, and therefore Eddington ratio.  To compare with other 
variability work they measure $SF^{(B)}$ of the sample, 
although they do not provide an error estimate.  

A different kind of AGN variability study was done by \cite{collier01}.  
Their sample consists of 12 Seyfert 1 galaxies with masses 
$\sim 10^{7} M_{\odot}$, observed every few days for up to several 
years.  The average structure function slope, measured 
over timescales of about 5 to 60 days, is listed as Sy1 in table 
\ref{results_comparison_table}.  Because they have many observations 
of the AGN they calculate the structure functions of each separately 
and see clear characteristic timescales of variability for 
each object.  The timescales 
differ widely between the AGN, ranging from 5 to 100 days.  
The average structure function slope of these Seyferts disagrees with 
our best value by 2.3$\sigma$.  However, the Seyferts' structure 
functions differ significantly from each other;  six out of the 
twelve structure function slopes are consistent with ours.  
Although we see no evidence for a turnover in the Palomar-QUEST ensemble 
structure function, it is likely that the individual objects in our 
sample may show structure function plateaus at various timescales.  
Such a detailed study of individual quasars requires more data than is 
available from the Palomar-QUEST Survey.

Table \ref{results_comparison_table} compares these different published 
results to the most comparable measurements from this work.  The QUEST1, 
SDSS + Plates, one of the SDSS + Spectra, and the SDSS Equatorial results 
do not take into account 
correlations between variability and other parameters such as luminosity 
or redshift;  these are compared to our unnormalized measurement.  One 
SDSS + Spectra measurement does deal with the parameter interdependencies;  
this is compared to our normalized result.  Our most robust result 
conservatively cuts out lower luminosity quasars;  there is no directly 
comparable result in the literature, but we include the measurement in the 
table as it is the most reliable.

\begin{table}
\begin{center}
\begin{tabular}{|l|l|l|}
\hline
Experiment & $SF$ Slope & Ref.\\\hline
\multicolumn{3}{|c|}{Unnormalized}\\\hline
QUEST2 & 0.357 $\pm$ 0.014 & (1)\\ 
QUEST1 & 0.47 $\pm$ 0.07 & (2)\\
SDSS + Plates & 0.30 $\pm$ 0.01 & (3)\\
SDSS + Spectra & 0.336 $\pm$ 0.033 & (4)\\
SDSS Equ. & 0.486 & (5)\\
Sy1 & 0.555 $\pm$ 0.030 & (6) \\\hline
\multicolumn{3}{|c|}{Normalized}\\\hline
QUEST2 & 0.392 $\pm$ 0.022 & (1) \\
SDSS + Spectra & 0.246 $\pm$ 0.008 & (4)\\\hline
\multicolumn{3}{|c|}{Normalized, Bright}\\\hline
QUEST2 & 0.432 $\pm$ 0.024 & (1)\\\hline
\end{tabular}
\end{center}
\caption{Comparison of experimental quasar structure function slopes. Reference (1) indicates this work.  (2): \cite{adam}. (3): \cite{devries05}. (4): \cite{vandenberk04}. (5): \cite{wilhite08}.  (6): \cite{collier01}.}
\label{results_comparison_table}
\end{table}

\cite{kawaguchi98} and \cite{hawkins02} present theoretical 
structure function slopes calculated from example disk instability, 
starburst, and microlensing models.  These slope predictions are 
shown in table \ref{theory_table}.  Our and most of the other results 
in table \ref{results_comparison_table} 
agree best with the disk instability prediction, 
although the models explored by those papers are by no means 
definitive;  more theoretical results with which to compare are 
needed.

\begin{table}
\begin{center}
\begin{tabular}{|l|l|l|}
\hline
Model Type & $SF$ Slope & Ref.\\\hline
Microlensing & 0.25 $\pm$ 0.03 & (1)\\
Disk Instability & 0.44 $\pm$ 0.03 & (2)\\
Starburst & 0.83 $\pm$ 0.08 & (2)\\\hline
\end{tabular}
\end{center}
\caption{Comparison of theoretical quasar structure function slopes. Reference (1) indicates \cite{hawkins02}.  (2): \cite{kawaguchi98}.}
\label{theory_table}
\end{table}

We see no evidence for asymmetry in quasar lightcurves over the 
timescales measured, as shown in figure \ref{sf_time_asymmetry}.  
\cite{devries05} calculated brightening and fading quasar structure 
functions over long timescales using several datasets and saw significant 
asymmetry on timescales of 2 to 3 years.  They suspect the 
asymmetry to be stronger than measured due to the dampening effects on 
the measurement by Malmquist bias.  Because our epochs 
are taken from the same dataset and the measurements at all epochs are 
treated identically, Malmquist bias will not affect the QUEST results.  
The timescales on which asymmetry was observed by \cite{devries05} 
lie in our longest timescale bins, which have large statistical and 
systematic errors.  Our null result is therefore not inconsistent 
with their asymmetry detection.

\subsection{Variability vs. Optical Luminosity}

The well-known trend that optically luminous quasars have fractional 
variability that is, on average, smaller than that of faint quasars is 
qualitatively consistent with the hypothesis that the variability is due 
to the cumulative effects of small discrete flares.  The details of 
the relationship between variability amplitude and quasar luminosity 
depend on the source of the variability.  For example, if the fluctuations 
were due to a Poissonian distribution of flares, each with identical 
timescale and energy, the following relationship would hold 
(\cite{cidfernandes00}):

\begin{equation} \frac{\Delta L}{\bar{L}} \propto \bar{L}^{-\delta} \label{deltaLequation}\end{equation}

where $\delta = \frac{1}{2}$.  This scenario is clearly too simplified 
to reflect the properties of quasar variability; however it illustrates 
the point that the relationship between variability amplitude and luminosity, 
and in particular the value of $\delta$, does probe 
the physics of the system.  Because $V$, defined in equation \ref{v_eq}, is 
an approximation of $\Delta m$, equation \ref{deltaLequation} is equivalent 
to the relation

\begin{equation} \log( V ) = K - \delta \cdot \log( L ) \label{sf_luminosity_eq} \end{equation}

where $K$ is a constant.  

We measure $\delta = 0.205 \pm 0.002$.  Clearly our results are inconsistent 
with the most basic Poissonian prediction.  \cite{vandenberk04} 
calculated $\delta$ using SDSS data to be 0.246 $\pm$ 0.005, which is 
inconsistent with our result at the level of 6$\sigma$.  Like 
ours, however, it is much shallower than the simple Poissonian prediction.

\subsection{Variability vs. Black Hole Mass}

\cite{wold07} have noted a positive correlation between quasar variability 
amplitude and black hole mass, using about 100 SDSS quasars seen in 
QUEST1 data.  Because they do not correct for correlations between 
quasar parameters, they note that their results with respect to mass may be 
confounded with trends with respect to time lag.  We confirm that the 
correlation is real, using much better statistics and more data at high 
masses.  The trend can be roughly described by the relation 
\[log(V) \propto \mu \times log(Mass).\]  
The best fit yields a reduced $\chi^{2}$ of 2.5 for $\mu = 0.13 \pm 0.01$.

\cite{collier01} notes a rough correlation between mass and characteristic 
variability timescale in their sample of 12 Seyfert 1 galaxies.  We see 
no such correlation in the quasars' average behavior over the 
timescales studied here.

\subsection{Variability vs. Eddington Ratio}

The previously observed dependences of variability on luminosity and 
mass imply that variability amplitude should decrease with Eddington ratio.  
This fact was discussed by \cite{wilhite08}, who measured the luminosity 
and mass relationships using repeated SDSS measurements of quasars 
in a 278 degree$^{2}$ equatorial strip and posited that the 
Eddington ratio may be the physically important parameter underlying 
both trends.  While they noted that their highest Eddington ratio 
quasars were indeed the least variable, they did not explore the shape 
of the trend as has been done in this work, due to the fact that their 
sample included 10 times fewer quasars than studied here.

We see flat average variability at both low and high Eddington 
ratios, with a transition around a ratio of 0.5.  
The time dependence of variability also changes with the 
quasar's Eddington ratio, as shown in figure \ref{sf_time_byer}.  
Quasars with low Eddington ratio tend 
to vary more on shorter timescales than those with high Eddington 
ratio, which vary more on longer timescales.  The behavior of the 
two groups is comparable around a rest frame time lag of 300 days.  
The variability mechanism 
therefore appears to change in quality between the two groups of objects, 
rather than simply quantity.  This could be consistent with 
fluctuations from different physical processes for the two groups.  Or, 
the Eddington ratio could reflect the same disk physics in each case, 
with lower Eddington ratios characteristic of disks with shorter variability 
timescales.  

\subsection{Variability vs. X-ray Emission}

We see an anticorrelation between optical variability amplitude 
and X-ray luminosity at the level of 2.9$\sigma$.  
Our result is not inconsistent with the overall increase in 
variability from X-ray non-detections to detections seen by 
\cite{vandenberk04} and \cite{adam}; the average variability 
amplitude of all of our X-ray detected quasars is higher than 
that of the remainder of the sample by more than $3\sigma$.  
\cite{adam} note that the few nonvariable quasars 
in their X-ray detected sample have higher average X-ray flux 
than the variable X-ray detected quasars, and suggest an anticorrelation 
between X-ray luminosity and optical variability amplitude.  Our 
results provide further evidence for such a relationship.

\subsection{Variability vs. Radio Emission}

There have been several reports of a positive correlation between optical 
variability amplitude and radio luminosity (e.g. \cite{adam}), although 
often the trend appears marginal (e.g. \cite{vandenberk04}).  
\cite{helfand01} studied roughly 200 radio-selected quasars and saw 
that the optical variability amplitude was fairly constant across the range of 
radio loudness except for a possible increase for the radio loudest 
objects.  We see no significant trend of optical variability amplitude 
with radio properties.  The radio fainter and brighter subsets and the radio 
non-detections all have indistinguishable variability amplitudes.  
Furthermore, the anticorrelation between optical variability amplitude 
and optical luminosity is present for both radio louder and quieter 
samples.  
The similarities in optical variability, both in amplitude and versus 
optical luminosity, between the radio 
samples undermines the idea that different mechanisms may be responsible 
for the variability in radio quiet and radio loud quasars, as 
conjectured by \cite{helfand01}.  

\subsection{Analogy with BHXBs \label{bhxb_section}}

Black hole X-ray binaries, like quasars, are systems containing a 
central black hole and an accretion disk.  BHXBs tend to be roughly 
$10^{7}$ times less massive than quasars, and their accretion disk 
emission is correspondingly faster and of higher energy.  X-ray 
observations of BHXBs over timescales of months 
can give insight into the behavior of quasars 
in optical and UV wavelengths over timescales that are too long to 
observe.  BHXB behavior may therefore be uniquely helpful in 
the interpretation of quasar observations by allowing the quasar 
data to be considered in terms of a longer timescale framework.

In black hole X-ray binaries three states are observed.  The low/hard 
state has Eddington ratios of a few percent or less, low X-ray luminosity, 
a hard spectrum, and radio jets.  
It is thought to be dominated by emission from an energetic corona that 
is related to the accretion disk in a manner not well understood, and 
which is likely causally associated with the formation of radio jets.  
The high/soft state has stronger accretion, with Eddington ratios of a 
few percent or more, 
higher X-ray luminosity, a soft spectrum, and no radio jets.  
Its emission is hypothesized to be dominated by the accretion disk.  
The intermediate state has some elements 
of the former states' spectral properties, Eddington ratios of either a 
few percent or tens of percent, sometimes exhibits jets, and has 
more dramatic X-ray variability than either the low/hard or high/soft 
states.  A thorough review of the X-ray variability of BHXBs can 
be found in \cite{vanderklis06}.

The smaller scale of BHXBs means that both disk 
and coronal emission lie in X-ray bands.  In quasar systems, the disk 
is further from the central engine and emits in the optical and UV 
range.  The corona, however, still emits in X-ray 
wavelengths.  Optical observations of quasars, then, may miss 
much of what corresponds to the low/hard X-ray state in BHXBs.  
Instead of observing disk-dominated emission and corona-dominated 
emission, we will observe primarily disk emission, but as modulated by 
other elements of the system.
%the influence of a corona that radiates X-rays.

We see a 
discrete shift in variability amplitude at intermediate Eddington 
ratio, implying that there may exist distinct phases in quasars that 
can be characterized by the Eddington ratio.  Do our low and high 
Eddington ratio quasars have properties that are consistent 
with the X-ray binaries' states?  Their average variabilities Log($V$), 
median X-ray loudnesses $\overline{\sigma_{ox}}$, and median radio 
loudnesses $\overline{R}$ are given 
in table \ref{eddington_table}.  The low Eddington ratio objects 
are on average more variable than the high Eddington 
ratio objects, as shown earlier.  
The median radio and X-ray properties are consistent between 
the two Eddington ratio groups.  The time dependencies of the two 
groups' optical variability are slightly different, as shown in figure 
\ref{sf_time_byer}, with the lower Eddington ratio group varying 
more strongly at shorter timescales, and less strongly at longer 
timescales.  These data best approximate the X-ray binary states 
if we compare our high Eddington sample with BHXBs in the high/soft 
state, and our low Eddington sample with BHXBs in the intermediate 
state.  This comparison matches the observed trends between Eddington 
ratio and average variability amplitude.  The stronger accretion 
implied by higher Eddington ratios may be associated with a larger 
accretion disk, and therefore longer variability timescales;  we 
indeed see longer variability timescales in the higher Eddington 
sample.  The dampened variability amplitude in the larger disk may be due to 
discrete flares making a smaller fractional contribution to the flux.  
Or, the heightened variability amplitude in the lower Eddington 
state may be due to interactions with other parts of the system such 
as the corona.

Eddington ratio is not clearly correlated with radio loudness, 
as might be expected by an analogy with BHXBs.  
The high/soft BHXB state does not exhibit radio jets, while the 
intermediate state has occasional radio outflows.  
Since the intermediate state has on average more radio emission, 
the radio detections should be more variable on average than radio 
nondetections.  The median radio loudness is indeed higher for 
the low Eddington ratio objects, as expected.  However, the 
distributions have significant overlap, as shown by the large 
median deviations.  A K-S comparison between the two sets of 
objects concludes that the radio loudnesses of the two groups 
has a probability $>99.9\%$ of belonging to 
different underlying distributions.  Therefore although we do not 
see a clearly defined trend, there is significant evidence that 
the two Eddington ratio groups have different radio properties.  
The low Eddington ratio subset is more radio loud than the high 
subset;  this is the sense expected from an analogy with the 
intermediate and high/soft BHXB states.

We also see no clear-cut correlation between Eddington ratio and X-ray 
loudness $\alpha_{ox}$.  If our lower Eddington ratio sample had 
stronger coronal emission, as is the case in the BHXB intermediate 
state, we would expect to see stronger X-ray emission in this sample.  
The median $\alpha_{ox}$ of the two groups does show this trend, 
although two distributions have significant overlap.  A K-S test 
between the two groups indicates that, despite the overlap, 
the $\alpha_{ox}$ of the two 
Eddington ratio subsets has $>99.9\%$ probability of stemming from 
different underlying distributions.  
It has been noted by \cite{vasudevan07} that $\alpha_{ox}$ may be a 
poor indicator of AGN X-ray to optical SED shape, as it stays roughly 
constant over a large range of observed spectral shapes.  Indeed, 
\cite{shemmer08} see only a very weak correlation between Eddington 
ratio and $\alpha_{ox}$ in a sample of 35 radio-quiet AGN, while 
measuring a much stronger correlation between Eddington ratio and 
$\Gamma$, the hard X-ray spectral slope.  This relationship between 
Eddington ratio and X-ray spectral hardness in quasars is of the 
same sense as the trend in BHXBs.  \cite{green08} observe a shallow 
but significant correlation between $\Gamma$ and $\alpha_{ox}$ in a 
sample of 1135 quasars;  this trend, while subtle, is analagous to behavior 
in BHXB X-ray spectra in the high/soft state.  A further link between 
AGN and BHXBs in X-ray observations is shown by \cite{vasudevan07}, 
who determine that at Eddington ratios above 0.1 the X-ray emission 
from the corona is around four times weaker with respect to the total 
bolometric emission than in AGN with lower Eddington ratios.  
This information reinforces our optical evidence for quasar 
states modulated by the Eddington ratio, where high ratio objects 
show characteristics similar to the BHXB high/soft state, which is 
dominated by disk rather than coronal emission.  

In summary, we see evidence for a distinct transition from a high 
accretion, larger disk scenario to one with stronger variability 
amplitude, shorter timescales, and less accretion.  
This behavior has parallels to the high/soft and intermediate 
states in BHXBs.  
We see some evidence for behavior from other parts of the system, 
such as a corona and jet, which may be associated with such a transition.  
The low and high Eddington ratio groups have significantly different 
radio and X-ray loudness properties.  The radio and X-ray distributions 
for the two groups significantly overlap, however, complicating an 
interpretation of the data.  

\begin{table}
\begin{center}
\begin{tabular}{|l|l|l|}
\hline
& $0 < E < 0.4$ & $0.6 < E < 1$\\\hline
Log($V$) & -0.922 $\pm$ 0.005 & -1.016 $\pm$ 0.006 \\
$\overline{\alpha_{ox}}$ & 1.29 $\pm$ 0.09 & 1.37 $\pm$ 0.09\\
$\overline{R}$ & 28 $\pm$ 22 & 21 $\pm$ 17\\\hline
\end{tabular}
\end{center}
\caption{Log($V$), median X-ray loudness $\overline{\alpha_{ox}}$, and median radio loudness $\overline{R}$ for quasars with low and high Eddington ratios $E$.}
\label{eddington_table}
\end{table}

\section{Conclusions}

We have studied the ensemble optical variability of roughly 23,000 
spectroscopically identified quasars using the Palomar-QUEST 
Survey.  The survey's data were carefully cleaned and recalibrated for 
this variability work.  The structure function was used to measure 
the quasar variability amplitude's dependence on time lag, and a 
power law exponent was fitted using several analysis methods.  
There is no evidence in our data for quasar lightcurve asymmetries.  
The variability amplitude was seen to decrease with increasing optical 
luminosity, a trend observed before although with a slightly different 
strength.  Black hole mass is positively correlated with optical variability 
amplitude, with a steady increase seen over several orders of magnitude in mass.  
The relationships between variability amplitude and both luminosity and mass 
imply a decrease in variability amplitude with increasing Eddington ratio.  We 
do see this trend;  in particular we observe steady variability 
levels at low and high Eddington ratios, with a transition around 
Eddington ratio of 0.5.  The increase in variability amplitude at low Eddington 
ratio is due to stronger fluctuations at timescales shorter than roughly 300 
days;  on timescales longer than 300 days the high Eddington ratio 
objects show stronger variability amplitude.  X-ray data for about 850 of the 
quasars show an anticorrelation between optical variability amplitude and 
X-ray luminosity.  Nearly 2,000 of the quasars have 
known radio fluxes;  they show no dependence of optical variability 
amplitude on radio luminosity or detection.  The discrete step nature 
of the optical variability amplitude vs. Eddington ratio results implies 
possible quasar states analogous to those in black hole X-ray binaries;  
our results indicate behavior similar to the binaries' intermediate 
state for low Eddington ratio quasars, and to the binaries' high/soft 
state for high Eddington ratio quasars.

\acknowledgements

We thank the Office of Science of the Department of Energy and the 
National Science Foundation for support.

{\it Facilities:} \facility{PO:1.2m}

\appendix

\section{Modelling of the Structure Function \label{modelling_appendix}}

The Palomar-QUEST time sampling is finite and nonuniform.  The 
structure function is therefore liable to include windowing effects due 
to the data cadence.  To examine this effect we simulate quasar 
lightcurves which yield a power-law structure function when sampled 
uniformly.  We then subject these lightcurves to our window function 
to study the effects of sampling.

\begin{figure}
\begin{center}
\plottwo{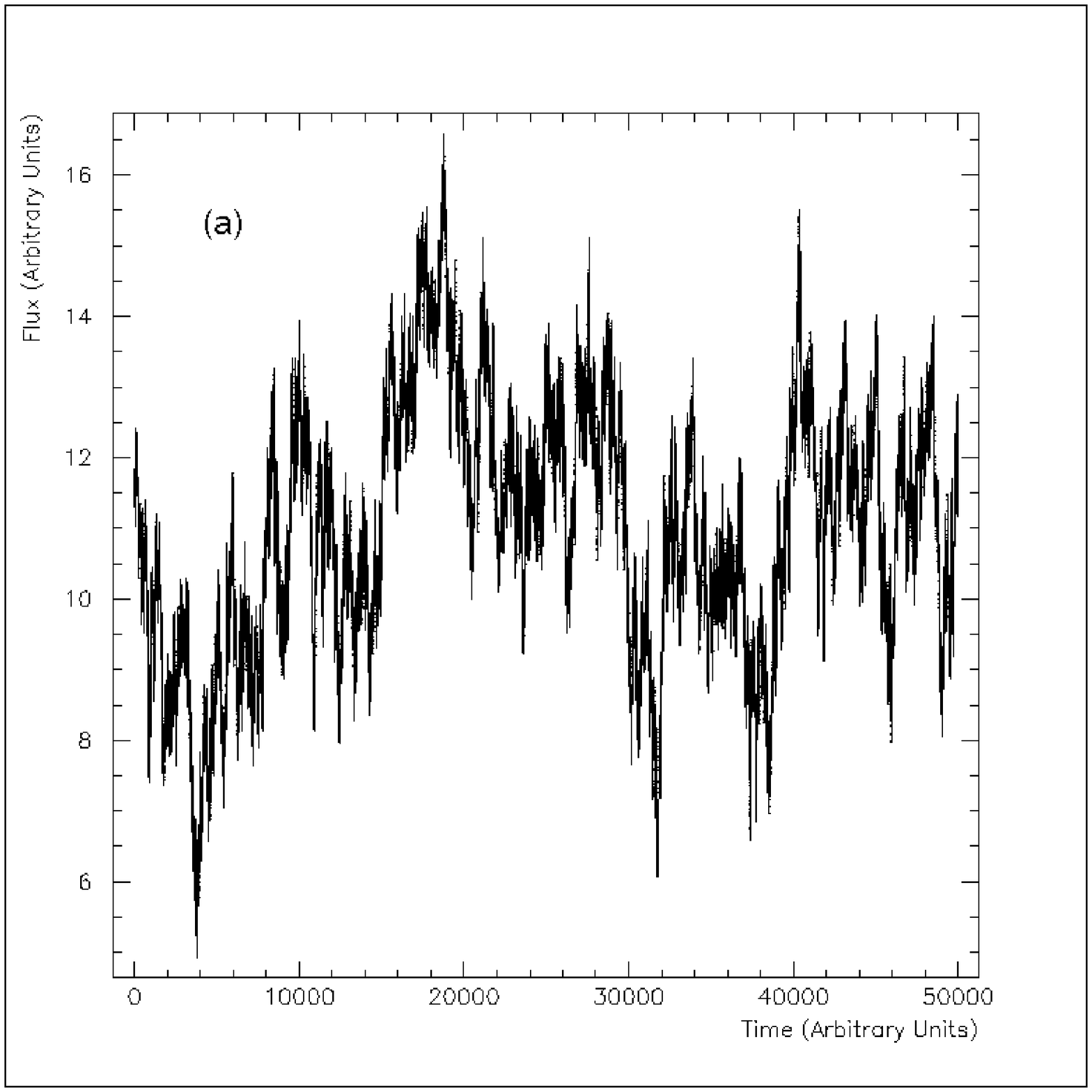}{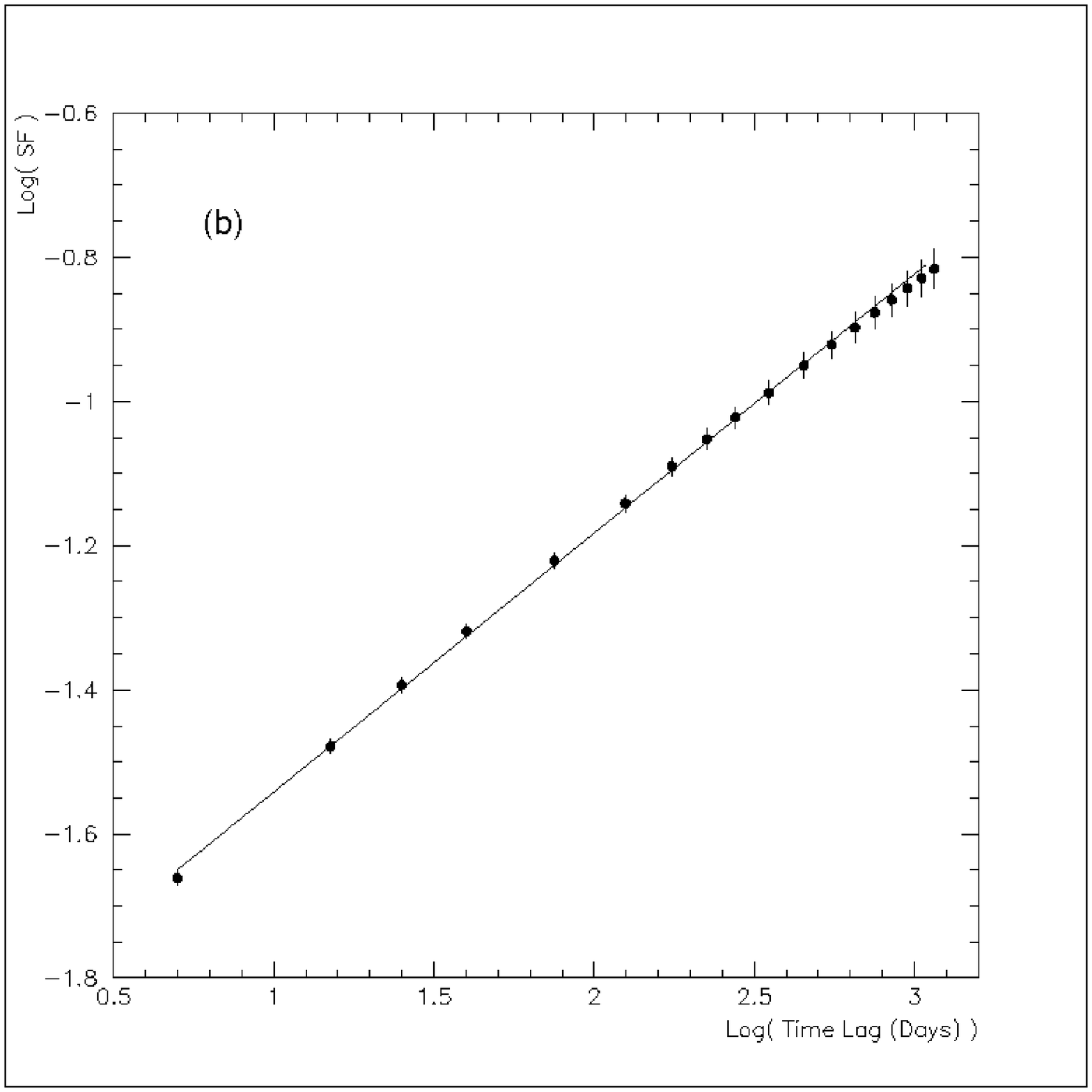}
\caption{(a): A sample simulated lightcurve with $\alpha = -1.71$.  (b): The structure function calculated from such simulations, with a fitted power law superimposed.  Plotted on a log-log scale.}
\label{sim_lc_sf}
\end{center}
\end{figure}

A structure function of the form $SF^{(A)} \propto t^{\beta}$ 
is equivalent to a power spectral distribution (PSD) of the form 
$PSD \propto f^{\alpha}$, 
where $\alpha = -2 \beta - 1$.  \cite{timmer95} present a method of 
computing light curves with power law spectra proportional to $f^{\alpha}$ 
through the careful choice of frequency amplitudes and phases.  We 
use their method to simulate light curves, and confirm the 
expected relationship between the frequency dependence of the lightcurves 
and the structure function slope.  An example evenly sampled, long timescale 
lightcurve and a structure function calculated from several such lightcurves 
are shown in figure \ref{sim_lc_sf}.  The lightcurve has a power spectral 
distribution $PSD \propto f^{-1.71}$, and the resulting structure function 
has logarithmic slope $\beta = 0.359 \pm 0.005$ which is consistent with the predicted 
value.  The error bars in the plot are determined from the distribution 
of multiple instantiations of the simulation.  The specific frequency 
dependence was chosen in order to approximately reproduce the slope of 
the experimental quasar structure function.

\begin{figure}
\begin{center}
\plotone{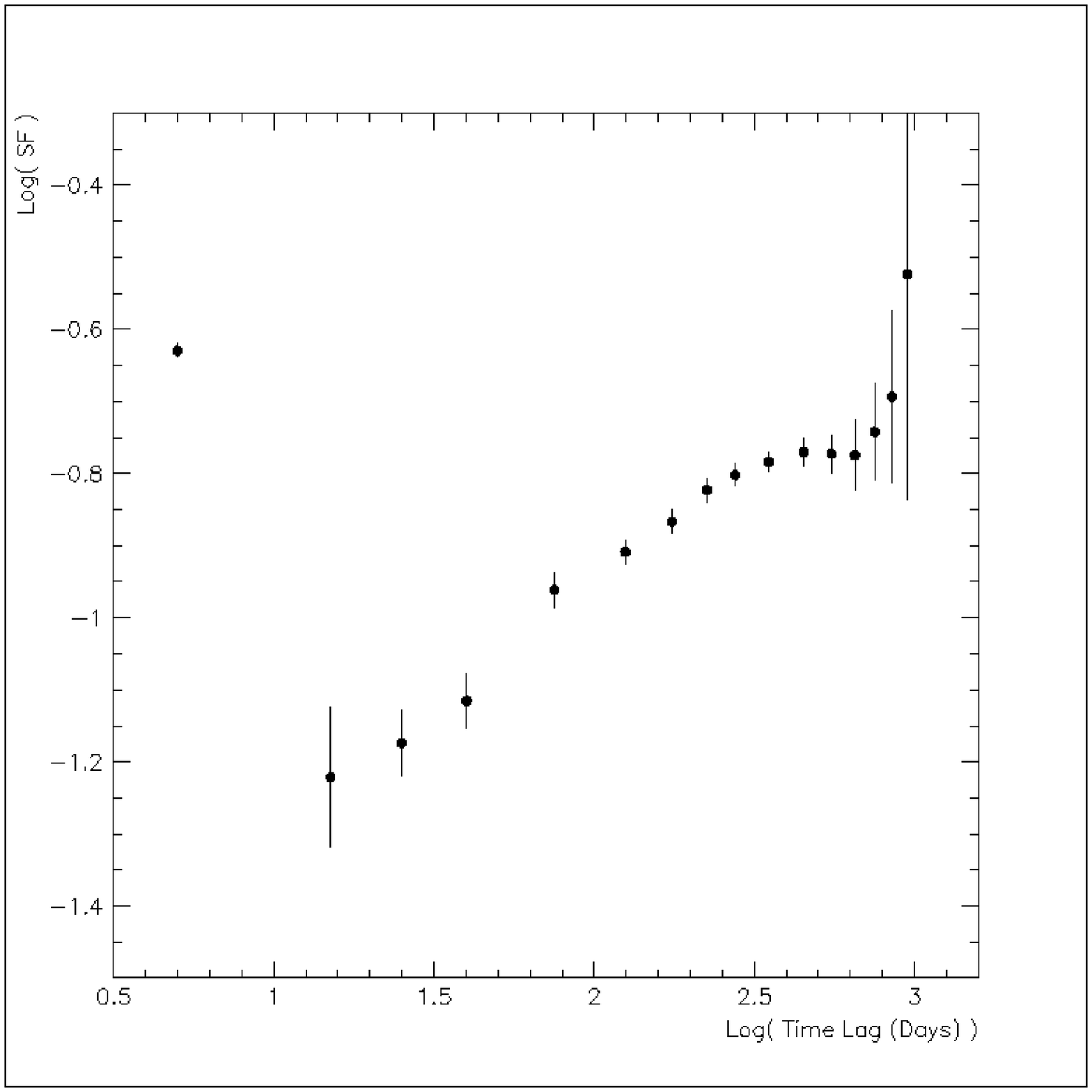}
\caption{The structure function calculated from simulated lightcurves and passed through the QUEST window function, with a fitted power law superimposed.  Plotted on a log-log scale.}
\label{windowed_sf}
\end{center}
\end{figure}

To study the effects of uneven data rates, we calculate the structure 
function from the simulated lightcurves using only data that align with 
our sampling cadence.  In particular, for each quasar in the sample we 
generate a simulated lightcurve, centered at the average flux value 
measured for that quasar.  We then replace each QUEST flux measurement 
with the value of the simulated lightcurve at the appropriate date, with 
added Gaussian noise on the order of our measurement errors.  In this way 
the times of our measurements are unchanged between the real and simulated 
data, allowing us to study our window function exactly.  The resulting 
simulated structure function is shown in figure \ref{windowed_sf}.  The 
error bars are determined by the spread between results from different 
subsets of the data, as in the analysis of the real data.  The amplitude 
of the lightcurves is set in order to match the resulting structure function 
plot's $y$ axis offset with the data.  In order to approximate the real data, 
the maximum lightcurve variations over the 3.5 year timescale of the survey 
were set to be roughly one magnitude.  Because the lightcurves have sharp 
peaks and our sampling is sparse, the true maxima were most often not 
observed.  The first data point is affected by obvious edge effects due to 
the fact that the minimum time difference sampled in the simulated lightcurve 
is one day while the real data extends to shorter timescales.  Note 
the flattening of the curve at high time lags where the window function is 
most influential.  The turnover in the experimental quasar structure function 
should therefore be attributed to a windowing effect rather than evidence 
for a maximum timescale of variability. 

The logarithmic slope of the linear region of the simulated, windowed structure function 
is measured to be $\beta = 0.341 \pm 0.029$, consistent with the value of 0.359 
$\pm$ 0.005 shown in figure \ref{sim_lc_sf} which was calculated using 
lightcurves with close, even sampling and the same frequency 
dependence.  The details of our data cadence therefore do not significantly 
affect the power law index of the structure function, given a power law 
frequency distribution of fluctuations.

The qualitative similarites in shape between our measured and simulated 
structure functions justifies the use of our chosen lightcurve model in 
examining the effects of uneven data sampling.  The errors on the windowed, 
simulated structure function, however, are larger than those in the real 
data.  This is because the spread in values of the structure function points 
is larger when using the simulated lightcurves.  This difference implies that 
the simulated lightcurves may not accurately reflect the true statistical 
distribution of the variability.  For the purposes of qualitatively 
understanding the effects of our data cadence, however, this parameterization 
is sufficient.

\end{document}